\documentclass[%
 reprint,
superscriptaddress,
 amsmath,amssymb,
 aps,
]{revtex4-2}
\usepackage{enumitem}
\usepackage{graphicx}% Include figure files
\usepackage{dcolumn}% Align table columns on decimal point
\usepackage{bm}% bold math
\usepackage[usenames,dvipsnames]{xcolor} % for edit macros
\usepackage[normalem]{ulem} % for strikes

\usepackage{xr} %start for cross ref
\makeatletter
\newcommand*{\addFileDependency}[1]{% argument=file name and extension
  \typeout{(#1)}
  \@addtofilelist{#1}
  \IfFileExists{#1}{}{\typeout{No file #1.}}
}
\makeatother

\newcommand*{\myexternaldocument}[1]{%
    \externaldocument{#1}%
    \addFileDependency{#1.tex}%
    \addFileDependency{#1.aux}%
}
\myexternaldocument{supp} %end for cross ref

\usepackage{xcolor}

\newif\ifJCCHighlightedChanges
\def\ifJCCHighlightedChanges{\iftrue}
\ifJCCHighlightedChanges
  
  \def\SJCC#1{{\color{blue}\sout{#1}}}

\else
  
  \def\SJCC#1{\relax}

\fi

\newif\ifURKHighlightedChanges
\def\ifURKHighlightedChanges{\iftrue}
\ifURKHighlightedChanges
  
  \def\SURK#1{{\color{red}\sout{#1}}}

\else
  
  \def\SURK#1{\relax}

\fi

\begin{document}

\preprint{APS/123-QED}

\title{Fimbriation alters bacterial colony morphogenesis in polymeric environments}

\author{Udayanidhi {Ramesh Kumar}}
\thanks{These authors contributed equally to this work.}
\affiliation{William A. Brookshire Department of Chemical and Biomolecular Engineering, University of Houston, Houston, TX, USA}

\author{Tanumoy Dhar}
\thanks{These authors contributed equally to this work.}
\affiliation{Division of Chemistry and Chemical Engineering, California Institute of Technology, Pasadena, CA, USA}

\author{Sebastian {Gonzalez La Corte}}
\thanks{Current address: Department of Molecular, Cellular, and Developmental Biology, Yale University, New Haven, CT, USA.}
\affiliation{Lewis-Sigler Institute for Integrative Genomics, Princeton University, Princeton, NJ, USA.}

\author{Sujit S. Datta}
\email{ssdatta@caltech.edu}
\affiliation{Division of Chemistry and Chemical Engineering, California Institute of Technology, Pasadena, CA, USA}

\author{Patrick C. Cirino}%
 \email{pccirino@uh.edu}
\affiliation{William A. Brookshire Department of Chemical and Biomolecular Engineering, University of Houston, Houston, TX, USA}

\author{Jacinta C. Conrad}%
 \email{jcconrad@uh.edu}
\affiliation{William A. Brookshire Department of Chemical and Biomolecular Engineering, University of Houston, Houston, TX, USA}

\date{\today}% It is always \today, today,

\begin{abstract}
In natural habitats, bacteria proliferate in macromolecule-laden fluids, forming multicellular colonies. The cells can alter the expression of surface appendages to adapt to their local niches, but how expression of these appendages affects the shape of a colony as it forms is not understood. Here, we show how a prominent bacterial appendage, nanofibers called type 1 fimbriae, alters the morphogenesis of a colony of \textit{Escherichia coli} proliferating in a polymer solution. Weakly fimbriated bacteria proliferate in extended, thick ``cables'' that are several cells wide, whereas strongly fimbriated bacteria proliferate in long, thin ``strings'' that are one cell wide. Using agent-based simulations, we show that the string structure arises from the competition between an entropic polymer-induced attraction, which favors lateral packing of cells, and the anisotropic repulsion imparted by the fimbriae, which opposes it everywhere except at the poles that appear after cells divide. Our study shows how the expression of bacterial appendages can be tuned to shape colonies, with potential implications for microbial functioning as well as for engineered living materials incorporating bacteria.

\end{abstract}

\maketitle

%\tableofcontents

\section{\label{section:intro}Introduction}

Bacteria often proliferate into multicellular colonies. The morphology of a colony can have critical consequences for its constituent cells: it controls how they access nutrients, signal to and compete with one another, and are exposed to external stressors such as antibiotics, phages, and host defenses~\cite{Chantratita2007morphology,Kragh2016,Melaugh2016,Secor2018,Simon-Soro2022,Secor2022,Kragh2023,Higgs2025motility}. Thus, understanding the mechanisms that shape the morphology of proliferating bacterial colonies is essential for understanding microbial processes and for the development of therapeutics and biomaterials.

One factor that may influence colony morphogenesis is interactions with extracellular polymers. In many natural settings, bacteria live and proliferate surrounded by polymeric macromolecules, such as glycosaminoglycans, glycoproteins, and collagen in host tissues \cite{Clark1988}; mucin, DNA, and F-actin in host secretions~\cite{Vasconcellos1994,Brandt1995,Fahy2010,McShane2021}; exopolymers in the environment~\cite{Decho2017}; and extracellular polymeric substances in bacterial aggregates~\cite{Jennings2021} and biofilms~\cite{flemming2010biofilm}. A particularly consequential example arises during infection by 
uropathogenic \textit{Escherichia coli} (UPEC), which invade bladder cells and proliferate into colonies encased by a polysaccharide-rich 
matrix~\cite{anderson2003intracellular}.

Recent work has shown that interactions with such extracellular polymers can dramatically shape proliferating colonies ~\cite{GonzalezLaCorte2025}: above a threshold polymer concentration, non-motile rod-shaped bacteria can proliferate into large-scale cable-like structures, driven by polymer-induced entropic depletion attractions between daughter cells and their hindered ability to diffusively separate after division. This prior study, however, did not examine bacteria expressing surface appendages. Among the most prevalent of these appendages are type 1 fimbriae, surface-anchored nanofibers~\cite{Pratt1998,Hancock2011, connell1996type,Brinton1965,Hospenthal2017} that are often displayed on cells in infections (including UPEC) and in biofilms~\cite{connell1996type,Pratt1998,Rodrigues2009,Staerk2015expression,AvalosVizcarra2016}. Hence, we ask: how does surface fimbriation shape the morphology of a bacterial colony proliferating in a polymeric environment? 

Here, we show that fimbriation strongly alters the morphology of colonies formed by non-motile \textit{E.\ coli} proliferating in polymer solutions. Experiments on bacteria with tunable fimbrial expression demonstrate that while fimbriae suppress depletion-driven aggregation entirely in the absence of growth, during proliferation they instead select colony shape: non- or weakly-fimbriated cells proliferate into cables that are multiple cells thick, whereas highly-fimbriated bacteria proliferate into ``strings'' that are only one cell wide. Combining experiments with theoretical modeling and agent-based simulations indicates that strings form because fimbriae coat the lateral surface of bacteria but are absent from the poles immediately after division; the steric repulsion they impart is therefore anisotropic, stabilizing cells against side-by-side but not end-to-end attraction. As a result, the degree of fimbriation selects the width of the emergent colony structure. Our work thus reveals how a single genetically tunable trait---expression of surface appendages---shapes proliferating bacterial colonies. It provides quantitative principles to predict and control these morphodynamics more broadly, with implications both for fundamental biological processes such as conjugation, contact inhibition, and short-range signaling and for practical applications in remediation of harmful biofilms and the design of antimicrobial treatments and engineered living materials.
 
\section{\label{section:results}Results}

\subsection{\label{results:nongrowth}Fimbriae suppress aggregation in non-proliferating bacterial suspensions}

\begin{figure}
\includegraphics[width=3.25in]{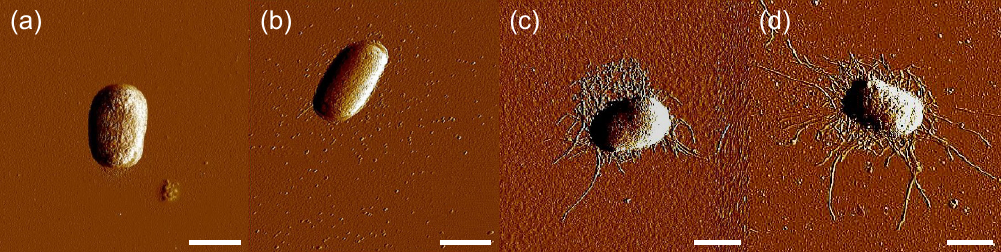}% 
\caption{\label{fig:AFM images of strains used in this study}Representative AFM images of \textit{E.\ coli} K-12 MG1655$\Delta$fliC transformed with plasmids (a) p(blank), (b) uninduced p(\textit{fim}), and p(\textit{fim}) induced with (c) 10 $\mu$M or (d) 50 $\mu$M IPTG for 4 h. Scale bar is 1 $\mu$m.}
\end{figure}

We construct a non-motile \textit{E.\ coli} mutant lacking expression of type 1 fimbriae and flagella. Many bacteria are able to proliferate in colonies without motility, and loss of motility can be an important strategy for bacterial persistence and biofilm formation \cite{Guttenplan2013lossofmotility,Mohaghegh2026motility}; thus, we focus on non-motile cells. In the reference \textit{E.\ coli} K-12 MG1655 strain, which lacks native expression of type 1 fimbriae~\cite{RameshKumar2024}, we delete the \textit{fliC} gene that encodes the flagellar filament to construct the flagella deletion mutant MG1655$\Delta$\textit{fliC}. MG1655$\Delta$\textit{fliC} is further transformed with a plasmid (that we call p(\textit{fim})) in which the \textit{fim} operon is under control of the chemically inducible \textit{tac} promoter; addition of IPTG to cell suspensions induces expression of type 1 fimbriae [Supplemental Material I]. A control plasmid without the \textit{fim} operon is called p(blank). To confirm that MG1655$\Delta$\textit{fliC}+p(\textit{fim}) (hereafter, p(\textit{fim})) produced fimbriae, we image the bacteria using atomic force microscopy (AFM). The MG1655$\Delta$\textit{fliC}+p(blank) (hereafter, p(blank)) strain produces nearly zero fimbriae [Fig.\ \ref{fig:AFM images of strains used in this study}(a)]. The p(\textit{fim}) strain expresses few fimbriae when no IPTG is added, indicating that the \textit{fim} operon is slightly leaky [Fig.\ \ref{fig:AFM images of strains used in this study}(b)]. When induced with higher concentrations of IPTG, however, this strain produces many fimbriae per bacterium [Fig.\ \ref{fig:AFM images of strains used in this study}(c), (d)]. p(\textit{fim}) bacteria induced with 10 $\mu$M IPTG bear approximately 19 fimbriae around the cell periphery with an average length of 0.6 $\mu$m, whereas those induced with 50 $\mu$M IPTG bear approximately 42 fimbriae per cell with an average length of 1 $\mu$m, as determined by counting the fimbriae located radially in the AFM images [Fig.\ \ref{fig:AFM images of strains used in this study}(c), (d)]. Thus, the number and length of type 1 fimbriae can be tuned by varying the IPTG concentration. 

% Figure 1: AFM images of strains used in this study

\begin{figure*}
\includegraphics{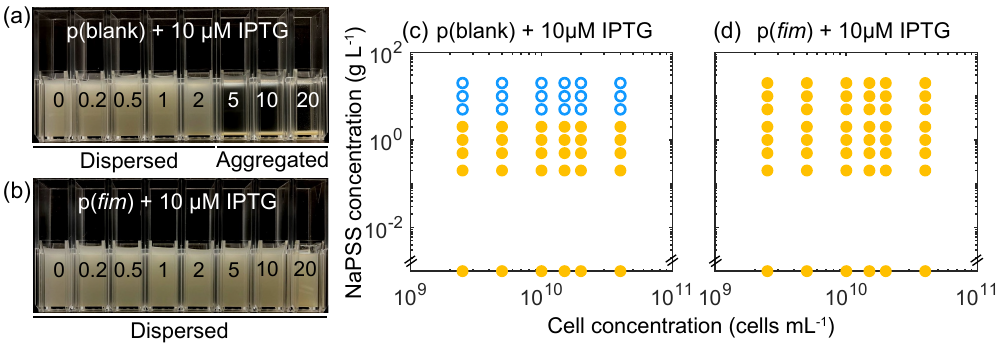}% 
\caption{\label{fig:sedimentation} Fimbriation inhibits depletion aggregation of \textit{E. coli} under non-growth conditions. (a, b) Cuvettes containing 4 $\times$ 10$^{10}$ cells mL$^{-1}$ of (a) p(blank) and (b) p(\textit{fim}) cells induced with 10 $\mu$M IPTG suspended in PBS with various concentrations of NaPSS (as labeled, in g L$^{-1}$) 4 h after addition of polymer. (c, d) State diagram for (c) p(blank) and (d) p(\textit{fim}) cells induced with 10 $\mu$M IPTG in PBS with NaPSS. Open blue and closed yellow circles respectively indicate suspensions in which cells form clusters or remain dispersed, determined visually 4 h after addition of polymer. p(blank) cells aggregate for [NaPSS] $\geq$ 5 g L$^{-1}$ ($U_\mathrm{ss}/k_B T \geq 143$). p(\textit{fim}) cells do not aggregate for NaPSS concentrations of up to 20 g L$^{-1}$.}
\end{figure*}

Non-adsorbing polymers added to suspensions of micron-sized bacteria induce entropic depletion attractions \cite{Eboigbodin2005,Schwarz-Linek2010EPL,Schwarz-Linek2012} to maximize the free volume available to the polymer. We first examine how depletion attractions affect the structure of bacteria under conditions in which the bacteria do not proliferate. We suspend p(blank) or p(\textit{fim}) bacteria induced with 10 $\mu$M IPTG in PBS, which does not support proliferation of bacteria, and add various amounts of sodium polystyrene sulfonate (NaPSS; $M_w = 70$ kDa), a negatively charged synthetic polyelectrolyte [Fig.\ \ref{fig:sedimentation}(a), (b)]. The range of the depletion attraction, set by the polymer radius of gyration $R_p = 22$ nm [Table SIII], is considerably shorter than the micron-long fimbriae. The magnitude of the polymer-induced depletion potential is greatest for rodlike bacteria when they aggregate side-by-side \cite{savenko2006phase} and proportional to the polymer concentration; we report the side-by-side potential $U_\mathrm{ss}$ normalized by the thermal energy $k_BT$ [Supplemental Material IV]. When the polymer concentration is less than or equal to $2$ g L$^{-1}$ ($U_\mathrm{ss}/k_B T \leq  57$), suspensions of non-fimbriated p(blank) bacteria are optically homogeneous after 4 h, indicating that the cells are uniformly distributed in a single phase. For NaPSS concentrations of 5 g L$^{-1}$ ($U_\mathrm{ss}/k_B T = 143$) or greater, however, the p(blank) suspension separates into dilute upper and concentrated lower phases, indicating that high concentrations of polymer induce bacteria to aggregate [Fig.\ \ref{fig:sedimentation}(b)].  Across the range of cell concentrations examined, suspensions of p(blank) bacteria separate into dilute and concentrated phases at sufficiently high concentrations of polymer. The boundary between the one- and two-phase regions is independent of cell concentration and lies between NaPSS concentrations of 2 and 5 g~L$^{-1}$ [Fig.\ \ref{fig:sedimentation}(c)]. The location of this state boundary is in near quantitative agreement with those reported in previous studies of non-flagellated \textit{E.\ coli} \cite{Schwarz-Linek2010EPL,Schwarz-Linek2010SM,Schwarz-Linek2012}; the strength of attraction required to observe phase separation in bacteria ($U_\mathrm{ss}/k_B T > 60$) is high compared to that required for phase separation of hard-sphere colloids (depletion attraction strength of $U/k_B T \approx 10$), but may reflect the need to overcome the electrostatic repulsion of negatively-charged bacteria. By contrast, suspensions of induced and hence fimbriated p(\textit{fim}) bacteria do not phase separate even in the presence of 20 g~L$^{-1}$ NaPSS for any cell concentration tested in this study [Fig.\ \ref{fig:sedimentation}(d)]. Similar state diagrams are observed upon addition of Ficoll, a neutral branched polymer (Supplemental Material VIII, Fig.\ S7). These findings suggest that fimbriae act as steric stabilizers that prevent depletion-driven aggregation under non-growth conditions and enable fimbriated cells to remain dispersed even in highly concentrated polymer solutions. 

% Figure 2: sedimentation

\begin{figure}
\includegraphics{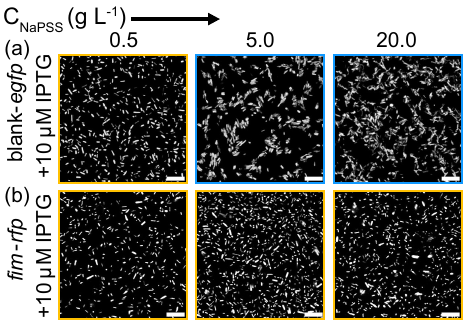}% 
\caption{\label{fig:nongrowth} (Color online) Confocal micrographs (grayscale) of (a) p(blank-\textit{egfp}) and (b) p(\textit{fim}-\textit{rfp}) cells induced with 10 $\mu$M IPTG in PBS containing NaPSS acquired 1 h after addition of polymer. Micrographs outlined in yellow or blue respectively indicate suspensions in which cells remain dispersed or form clusters. p(blank-\textit{egfp}) cells induced with 10 $\mu$M IPTG remain dispersed at low NaPSS concentration of 0.5 g L$^{-1}$ ($U_\mathrm{ss}/k_B T = 14$), however, aggregate side-by-side at high NaPSS concentration of 20 g L$^{-1}$ ($U_\mathrm{ss}/k_B T = 571$). By contrast, p(\textit{fim}-\textit{rfp}) cells induced with 10 $\mu$M IPTG remain dispersed even at 20 g L$^{-1}$ NaPSS. Scale bar is 10 $\mu$m.}
\end{figure}

In the absence of fimbriae side-by-side aggregation of cells is favored over end-to-end aggregation, because the ratio of the side-by-side $U_\mathrm{ss}$ to end-to-end $U_\mathrm{ee}$ depletion potentials \cite{savenko2006phase}  $U_\mathrm{ss}/U_\mathrm{ee} \sim 8$ for all polymer concentrations. Fimbriated bacteria, by contrast, are not expected to make contact in either configuration because the length of fimbriae exceeds the range of the depletion attraction. To test this idea, we image bacteria suspended in solutions of NaPSS in PBS. Genes for constitutive expression of the enhanced green fluorescent protein (\textit{egfp}) and red fluorescent protein (\textit{rfp}) are incorporated into the p(blank) and p(\textit{fim}) plasmids, respectively, to enable imaging of non-fimbriated and fimbriated MG1655$\Delta$\textit{fliC} bacteria using confocal microscopy. Without added polymer, p(blank-\textit{\textit{egfp}}) cells remain dispersed in solution [Fig.\ S6(a) in Supplemental Material]. When the concentration of NaPSS is 5 g~L$^{-1}$ ($U_\mathrm{ss}/k_BT = 143$), however, they align side-by-side to form clusters [Fig.\ \ref{fig:nongrowth}(a)] consistent with prior studies of depletion-driven aggregation in rod-shaped bacteria \cite{Eboigbodin2005,Schwarz-Linek2010EPL,Schwarz-Linek2010SM,Dorken2012,Schwarz-Linek2012,Secor2018,Peters2021,Secor2022}. Further increase in polymer concentration leads to increases in both the number and size of clusters [Fig.\ \ref{fig:nongrowth}(a)]. By contrast, p(\textit{fim}-\textit{rfp}) cells induced with 10 $\mu$M IPTG remain dispersed for NaPSS concentrations of up to 20 g L$^{-1}$ [Fig.\ \ref{fig:nongrowth}(b)]. 

The results presented in Figs.\ \ref{fig:sedimentation} and \ref{fig:nongrowth} reveal that  fimbriae abolish depletion-driven aggregation. Whereas the aggregation boundary for non-fimbriated bacteria lies between 2 and 5 g L$^{-1}$ NaPSS (Fig.\ \ref{fig:sedimentation} and Fig.\ S6), fimbriated bacteria show no aggregation even at polymer concentrations of up to 20 g L$^{-1}$. 

% Figure 3: nongrowth

% Figure 4: growth

\begin{figure*}
\includegraphics{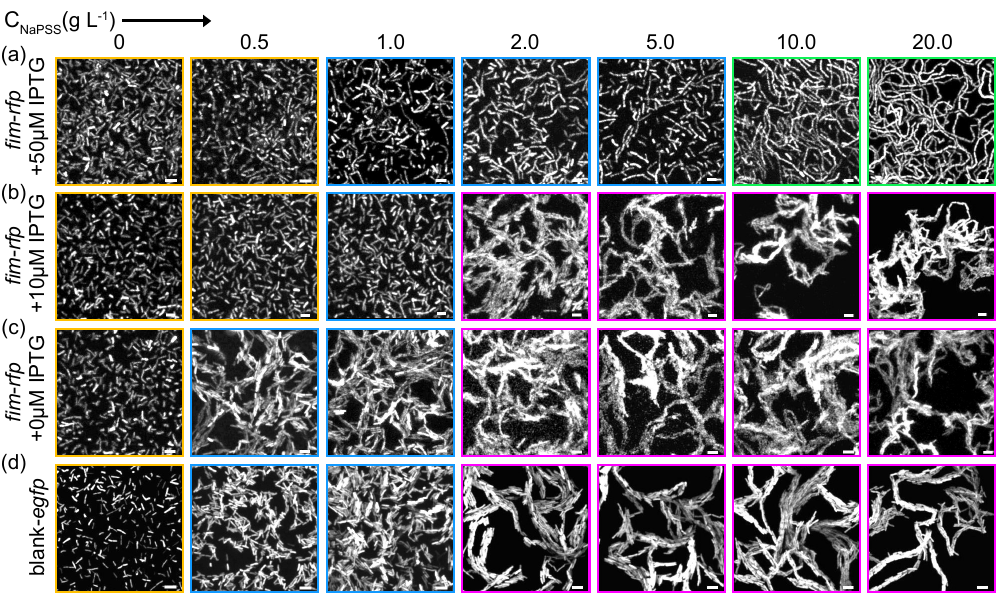}% 
\caption{\label{fig:growth} (Color online) Morphogenesis of bacterial networks depends on fimbriation. (a -- d) 2-D projections of 3-D confocal micrographs (grayscale) of p(\textit{fim}-\textit{rfp}) cells induced with (a) 50 $\mu$M IPTG, (b) 10 $\mu$M IPTG, and (c) 0 $\mu$M IPTG and (d) p(blank-\textit{egfp}) after 5 h of proliferation in LB medium containing NaPSS at various concentrations indicated in the figure. Micrographs outlined in yellow, blue, magenta, or green respectively indicate dispersed cells, clusters, cables, and strings. The end-to-end depletion potentials ($U_\mathrm{ee}/k_B T)$ corresponding to [NaPSS] = 0, 0.5, 1.0, 2.0, 5.0, 10.0, and 20.0 g L$^{-1}$ are 0, 1.7, 3.4, 6.9, 17, 34, and 69, respectively. Morphologies of cells proliferating in nutrient-rich NaPSS solution can be altered by tuning the degree of fimbriation and strength of depletion attractions. Scale bar is 5 $\mu$m.}
\end{figure*}

\subsection{\label{results:growth}Fimbriae alter colony morphogenesis in nutrient-rich polymer solutions}

% The boundary between clusters and cables does not coincide with the polymer overlap concentration $c^*$.

Next, we image p(blank-\textit{egfp}) and p(\textit{fim}-\textit{rfp}) bacteria as they proliferate in nutrient-rich LB media containing NaPSS using confocal microscopy 
\footnote{The polymer radius in LB media, $R_p = 13$ nm [Table SIII], is slightly smaller than the value of 22 nm determined in PBS.}. When cells proliferate, they grow end-to-end; although $U_\mathrm{ss} > U_\mathrm{ee}$, because the cells are end-to-end after division we take $U_\mathrm{ee}$ as the characteristic depletion strength hereafter. In the absence of polymer, p(blank-\textit{egfp}) cells diffuse away after division to form a random dispersion [Fig.\ \ref{fig:growth}(d); the classification heuristic is given in Section \ref{methods:classify}]. At low concentrations of polymer ($\leq  1$ g L$^{-1}$, end-to-end depletion potential \cite{savenko2006phase} $U_\mathrm{ee}/k_B T \leq 3.4$), p(blank-\textit{egfp}) bacteria proliferate into clusters in which the cells are aligned both side-by-side and end-to-end. For polymer concentrations greater than $2$ g L$^{-1}$ ($U_\mathrm{ee}/k_BT \geq 6.9$), however, the bacteria proliferate into long, multicellular cable-like networks, as previously reported \cite{GonzalezLaCorte2025}. The proliferating non-fimbriated bacteria remain attached end-to-end after division but are unable to diffuse apart due to the polymer-induced depletion attractions \cite{GonzalezLaCorte2025}. After the next division the bacteria form a metastable, interdigitated two-by-two cluster in which the daughter cells are retained end-to-end; this cluster nucleates the thick multicellular cables in which bacteria are nematically aligned \cite{GonzalezLaCorte2025}. Uninduced p(\textit{fim}-\textit{rfp}) bacteria exhibit similar states as p(blank-\textit{egfp}): they are dispersed in the absence of polymer, form clusters with side-by-side and end-to-end attractions below polymer concentrations of $1$ g L$^{-1}$ ($U_\mathrm{ee}/k_BT \leq 3.4$), and proliferate into cables at large polymer concentrations ($\geq 2$ g L$^{-1}$, $U_\mathrm{ee}/k_B T \geq 6.9$) [Fig.\ \ref{fig:growth}(c)]. Thus, the location of the boundary between the cluster state and the cable state does not differ between p(blank-\textit{egfp}) and uninduced p(\textit{fim}-\textit{rfp}) strains. This result shows that the weak expression of fimbriae in the uninduced p(\textit{fim}-\textit{rfp}) bacteria [Fig.\ \ref{fig:AFM images of strains used in this study}(b)] does not quantitatively alter the state boundaries for morphogenesis. 

To generate fimbriae, we induce expression of the \textit{fim} operon by adding IPTG. Fimbriated p(\textit{fim}-\textit{rfp}) cells induced with 10 $\mu$M IPTG proliferate as a random dispersion when the concentration of added polymer is low ($\leq 0.5$ g L$^{-1}$, $U_\mathrm{ee}/k_BT = 1.7$) [Fig.\ \ref{fig:growth}(a,b)]. The dispersed state persists at higher concentration of polymer for these fimbriated bacteria than for the nonfimbriated p(blank-\textit{egfp}) or very weakly fimbriated p(\textit{fim}-\textit{rfp}) + 0 $\mu$M IPTG strains, indicating that fimbriae stabilize the cells against aggregation in the presence of weak depletion attractions. When the polymer concentration is 1 g L$^{-1}$ ($U_\mathrm{ee}/k_BT = 3.4$), the bacteria form clusters with end-to-end but not side-by-side attractions, indicating that fimbriae partially stabilize the bacteria against lateral aggregation. At high concentrations of polymer, the degree of fimbriation markedly alters colony morphogenesis. Fimbriated p(\textit{fim}-\textit{rfp}) cells induced with 10 $\mu$M IPTG form cable-like networks with end-to-end and side-by-side attractions for polymer concentrations greater than $2$ g L$^{-1}$ ($U_\mathrm{ee}/k_B T \geq~6.9$) [Fig.\ \ref{fig:growth}(b)]. The cables are similar to those formed by the p(blank-\textit{egfp}) strain and the uninduced p(\textit{fim}-\textit{rfp}) strain. In sharp contrast, for polymer concentrations exceeding 10 g L$^{-1}$ ($U_\mathrm{ee}/k_BT \geq 34$), highly fimbriated bacteria (induced with 50 $\mu$M IPTG) are held end-to-end as they proliferate but do not aggregate side-by-side as in cables---ultimately producing ``string''-like morphologies that are only one cell wide [Fig.\ \ref{fig:growth}(a)]. Thus, highly-fimbriated cells appear to feel only end-to-end but not side-by-side depletion attractions. Since fimbriae do not appear to be expressed at the poles of the cell body immediately after division (i.e., at the conjunction of two dividing cells) [Fig.\ S4 in Supplemental Material], these results suggest that the abundant fimbriae on their lateral surfaces enable bacteria to resist side-to-side aggregation even at high concentrations of polymer, but they nonetheless aggregate end-to-end after division because their poles lack fimbriae. This resistance to lateral but not end-to-end aggregation leads to thin strings, rather than the thick cables observed at lower IPTG induction levels (10 $\mu$M, Fig.\ \ref{fig:growth}(c)). 

% These bacteria form clusters when the polymer concentration is 0.5 g L$^{-1}$ ($U_\mathrm{ee}/k_BT = 1.7$) but the uninduced p(\textit{fim}) bacteria do not; this result confirms that the p(\textit{fim}) induced at 10 $\mu$M IPTG are fimbriated during proliferation. 

The p(\textit{fim}) bacteria induced at 10 $\mu$M IPTG form cables during proliferation (Fig.\ \ref{fig:growth}(b)) but do not aggregate under non-proliferating conditions (Fig.\ \ref{fig:sedimentation}). These bacteria are fimbriated during proliferation: they form clusters for NaPSS concentrations of 0.5 g L$^{-1}$ ($U_\mathrm{ee}/k_BT = 1.7$) whereas the uninduced p(\text{fim}) bacteria do not. Nonetheless, they proliferate into cables when grown in solutions of high polymer concentration. To understand the origins of these cables, we image bacteria induced at 10 $\mu$M IPTG over time during colony morphogenesis. We find that these bacteria form strings at early times but cables on long time scales (Fig.\ S9).  AFM measurements on cells grown in LB without polymer, under otherwise matched growth conditions, show that the distribution of fimbriae per cell broadens with time, with approximately 5\% of cells bearing no fimbriae at 90 min and approximately 40\% bearing no fimbriae at 240 min (Fig.\ S10). Thus, modest fimbriation does not preclude string formation, but cables form when some (not necessarily all) cells lose fimbriation; once a few non-fimbriated cells arise within a string, depletion attractions can draw them side-by-side with their neighbors and nucleate cables. The transition from strings to cables over time is consistent with the stochastic nature of protein expression from the (LacI-repressed) \textit{lac} promoter when induced at ``low'' IPTG concentrations (here, 10 $\mu$M) \cite{10.1021/acs.biochem.9b01101}, due in part to expression of \textit{lacY} (encoding an active transporter for IPTG) gradually turning off \cite{BHOGALE2022183}.

Together, our experiments suggest that fimbriae alter the morphogenesis of bacterial networks in polymer solutions. Highly fimbriated bacteria proliferate end-to-end but do not pack in the energetically favorable side-by-side configuration. Thus, fimbriae on the lateral cell surface must impede side-by-side aggregation. Because fimbriae are absent at the poles immediately after division, however, the fimbriated bacteria remain attached end-to-end due to polymer-induced depletion attractions at their poles. Subsequent divisions generate a one-cell-wide string of bacteria that is built exclusively from the weaker end-to-end depletion attraction rather than the stronger side-by-side attraction. Notably, it is division during proliferation that generates the attachment point at the poles; under the non-growth conditions, the poles are fimbriated and bacteria cannot attach end-to-end.

\begin{figure*}
\includegraphics[width=\textwidth]{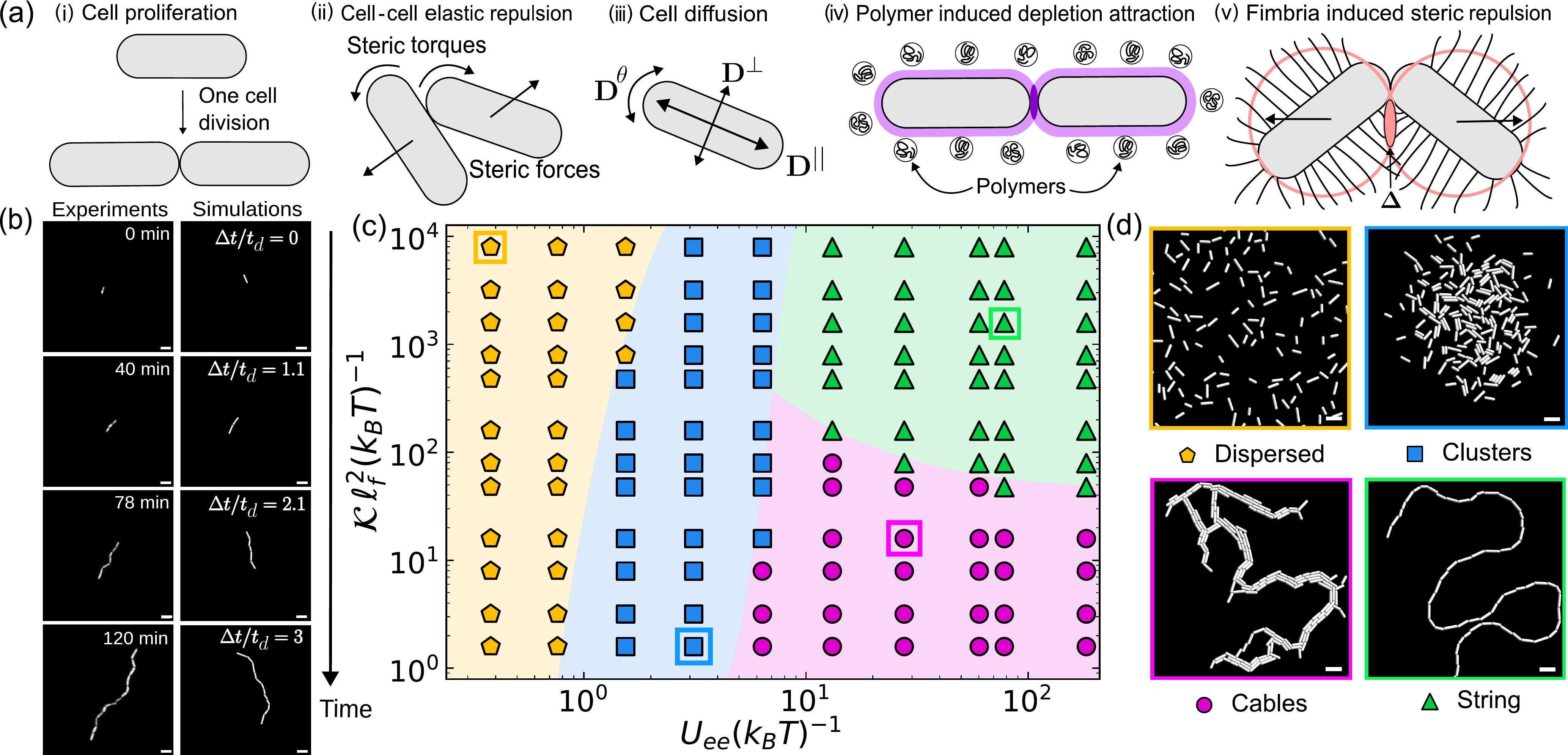}% 
\caption{\label{fig:simulations} (Color online) Agent-based simulations recapitulate colony morphogenesis in polymeric fluids. (a) Schematic description of the five key biophysical features underlying the simulations: bacterial cells (i) continually grow and divide, (ii) do not overlap, (iii) undergo longitudinal, transverse, and rotational diffusion with diffusion coefficients $\mathbf{D}^{||}$, $\mathbf{D}^{\perp}$, and $\mathbf{D}^{\theta}$, respectively, (iv) are held together by polymer-induced depletion attraction, and (v) have lateral steric repulsion induced by fimbriae that overlap by an amount $\Delta$. (b) Time series of a string formed by fimbriated cells proliferating in a polymer solution in experiments (left) and simulations (right). Experiments show p(\textit{fim}-\textit{rfp}) cells induced with 50 $\mu$M IPTG proliferating in LB medium containing 20~g~L$^{-1}$ NaPSS; simulations are conducted at $U_\mathrm{ee}/k_BT = 78$ and $\mathcal{K}\ell_f^2/k_BT = 1586$, at the elapsed times indicated normalized by the doubling time $t_d$. Scale bars show 5~$\mu$m. (c) Morphological state diagram spanned by the two dimensionless parameters $U_\mathrm{ee}/k_BT$ and  $\mathcal{K}\ell_f^2/k_BT$, where the former compares the magnitude of the depletion potential to thermal energy and the latter compares the maximal energy required to compress the fimbrial layer to the thermal energy. Yellow pentagons, blue squares, magenta circles, and green triangles indicate colonies that form random dispersions, small clusters, cables, and strings, respectively; the color shading is a guide to the eye. (d) Snapshots of representative colony morphologies corresponding to the points indicated in (c) for each of these four states: dispersed ($U_\mathrm{ee}/k_BT = 0.4,~\mathcal{K}\ell_f^2/k_BT = 7931.5$, $\Delta t/t_d\sim 7.5$), clusters 
($U_\mathrm{ee}/k_BT = 3.1,~\mathcal{K}\ell_f^2/k_BT = 1.6$, $\Delta t/t_d\sim 7.5$), cables ($U_\mathrm{ee}/k_BT = 27.7,~\mathcal{K}\ell_f^2/k_BT = 15.9$, $\Delta t/t_d\sim 7.5$), and strings ($U_\mathrm{ee}/k_BT = 78.2,~\mathcal{K}\ell_f^2/k_BT = 1586.3$, $\Delta t/t_d \sim 6$). Scale bars show 5~$\mu$m. }
\end{figure*}

\subsection{\label{results:model} Anisotropic steric repulsions from lateral fimbriae cause string formation}

To test whether steric repulsion from fimbriae is sufficient to produce strings, we perform two-dimensional agent-based Brownian dynamics simulations, building directly on those developed to describe cable formation~\cite{GonzalezLaCorte2025} as detailed in the Supplemental Material. Each simulation begins with a dilute dispersion of cells, just as in the experiments, each of which is modeled as a spherocylinder that proliferates in a polymer solution with a prescribed polymer size $R_p$, viscosity $\eta$, and osmotic pressure $\Pi^\mathrm{osm}$. The simulations incorporate five key biophysical features, schematized in Fig.~\ref{fig:simulations}(a) and described theoretically in the Supplemental Material: 
\begin{enumerate}[label=\roman*.]
\item Cells continually elongate exponentially, separating into two daughter cells after a doubling time $t_d$.
\item Adjacent cells have a short-ranged elastic repulsion that prevents their bodies from overlapping.
\item Cells undergo both translational and 
rotational diffusion in the polymer solution.
\item Cells are subject to the polymer-induced end-to-end depletion attraction given by a pairwise interaction potential of magnitude $U_\mathrm{ee} = \Pi^\mathrm{osm} V_\mathrm{ev}$, where the excluded volume $V_\mathrm{ev}$ is set by the polymer size and the cell sizes and orientations~\cite{Asakura1958,savenko2006phase}.
\item Cells are surrounded by a soft exclusion region representing the fimbriated cell surface, which extends beyond the lateral surface but vanishes at the poles, reflecting the absence of fimbriae there immediately after division [Fig.~S4].
\end{enumerate}
This fifth feature represents a new addition to our previous model of cable formation~\cite{GonzalezLaCorte2025}, which did not consider fimbriae.

Each simulation can then be parameterized by three dimensionless parameters: $U_\mathrm{ee}/k_BT$, which compares the magnitude of the depletion potential for cells arranged end-to-end after doubling $U_\mathrm{ee}$ to the thermal energy $k_BT$; $t_\mathrm{rot}/t_d$, which compares the characteristic timescale for cells to rotationally diffuse $t_\mathrm{rot}$ to their doubling time $t_d$; and a new parameter $\mathcal{K}\ell_f^2/k_BT$, which compares the maximal energy required to compress the fimbrial layer, of effective stiffness $\mathcal{K}$ and characteristic lateral maximal reach $\ell_f$, to the thermal energy. For simplicity, we hold $t_\mathrm{rot}/t_d$ fixed throughout; our prior work \cite{GonzalezLaCorte2025} showed that varying this parameter does not strongly alter colony morphologies, as we further verify in Fig.~S12.

When fimbriation is weak ($\mathcal{K}\ell_f^2/k_BT \sim 10^{0}$), the simulations reproduce our prior results on non-fimbriated cells~\cite{GonzalezLaCorte2025}: in sufficiently concentrated polymer solutions ($U_\mathrm{ee}/k_BT \gtrsim 4$), cells quickly lock in side-by-side after division and continue to proliferate end-to-end, forming cables that are multiple cells thick. Increasing the level of fimbriation changes these colony morphodynamics entirely. At comparable depletion strength but $\mathcal{K}\ell_f^2/k_BT \sim 10^{3}$, for example, side-by-side packing is suppressed: cells proliferate exclusively end-to-end, forming a single-cell-wide string that elongates and eventually buckles [Fig.~\ref{fig:simulations}(b)], just as in the experiments.

This selection of end-to-end over side-by-side attachment between cells reflects a competition between two opposing anisotropies. Because fimbriae coat the sides of a cell but not its newly formed poles (Fig.\ S4), two cells approaching side-by-side must compress the fimbrial layer along their entire length, whereas two newly-divided cells are held pole-to-pole before their fimbriae come into contact. Conversely, the depletion attraction induced by extracellular polymers is anisotropic in the opposite sense, holding cells together stronger when they are side-by-side versus end-to-end \cite{savenko2006phase,GonzalezLaCorte2025}. Thus, depletion always favors lateral packing of cells, whereas fimbriae always oppose it. We therefore expect strings to form when the depletion attraction is strong enough to hold daughter cells together at their poles \emph{and} the fimbrial repulsion is strong enough to prevent them from subsequently packing side-by-side---that is, at sufficiently large values of both $U_\mathrm{ee}/k_BT$ \emph{and} $\mathcal{K}\ell_f^2/k_BT$.

To test this expectation, we map the morphological state diagram spanned by these two parameters, varying the value of each over several decades. Our results, summarized in Fig.~\ref{fig:simulations}(c), confirm this expectation and recapitulate all four colony morphologies observed experimentally, as shown by the representative examples in Fig.~\ref{fig:simulations}(d). At small polymer concentrations ($U_\mathrm{ee}/k_BT \lesssim 0.8$), cells grow, divide, and diffuse freely, remaining dispersed regardless of fimbriation [yellow pentagons]. At weak levels of fimbriation ($\mathcal{K}\ell_f^2/k_BT \lesssim 40$), increasing polymer-induced depletion (increasing $U_\mathrm{ee}/k_BT$) produces small clusters of cells [blue squares], eventually leading to cables when $U_\mathrm{ee}/k_BT\gtrsim6$ [magenta circles]---just as in the experiments [Fig.\ \ref{fig:growth}(b-d)]. With sufficiently strong fimbriation ($\mathcal{K}\ell_f^2/k_BT \gtrsim 10^{2}$), however, cells grow into strings, not cables, over the same range of depletion strengths [green triangles]---again, just as in the experiments [Fig.\ \ref{fig:growth}(a)]. Thus, despite the simplifications made in the model, our simulations  recapitulate the influence of fimbriation on colony morphogenesis that we observe experimentally. Moreover, they show that the competition between two opposing anisotropies is sufficient to select one colony morphology over the other: the depletion attraction, which favors side-by-side packing, against the fimbrial repulsion, which opposes it everywhere but at the poles newly created by division.

\section{Discussion}

Multicellular morphologies reminiscent of strings have been observed in hydrostatic biofilms at air-liquid interfaces \cite{Puri2024}, in the human oral microbiome \cite{MarkWelch2016}, and in liquid crystal media \cite{GonzalezLaCore2026chains} among other settings. Overexpression of a surface adhesin, type 1 fimbriae, alters morphogenesis in \textit{E.\ coli} proliferating in polymer solutions. Whereas in the absence of fimbriae bacteria proliferate in cables, as previously observed~\cite{GonzalezLaCorte2025}, at high levels of fimbrial expression bacteria instead form strings that are one cell wide. By combining experiments, theory, and simulations, we show that the string morphology arises from the competition between two anisotropic interactions: depletion attractions favor packing of rodlike bacteria side-by-side, whereas fimbrial repulsions oppose clustering of bacteria everywhere but at the bare poles of cells that are created after bacteria divide. Even though the end-to-end depletion attraction is approximately eight times weaker than the side-by-side attraction \cite{savenko2006phase}, it is this weaker attraction that drives string formation in highly fimbriated bacteria because of the strong lateral repulsion imparted by fimbriae.

The model underlying our simulations captures the essential biophysical mechanisms underlying string formation in polymer solutions. However, it necessarily involves simplifications. For example, we assumed that fimbriae are not present at the cell poles---that is, fimbrial regrowth at the poles after cell division is taken to be much slower than the doubling time $t_d$ \cite{Puorger2011fimAfolding}. Rather than treating each appendage individually, we coarse-grained the fimbrial layer into a single soft exclusion region centered on the cell, resulting in central forces that exert no torque. Moreover, because the stiffness of the fimbrial layer has not been measured independently, a quantitative comparison between the simulations and experiments is out of scope of the present work. Incorporating these complexities into the model, and directly measuring the mechanical response of the fimbrial layer to better parameterize the simulations, present opportunities for future research. 

The ability to generate thin filaments of single-cell width from sequential divisions has implications for numerous biological processes. First, bacteria proliferating in a linear string whose length $L$ grows proportional to the number of bacteria $N$, i.e.\ $L \sim N$, explore far more space than bacteria proliferating in a sphere (for which the radius $R \sim N^{1/3}$) or even a thicker cable \cite{GonzalezLaCorte2025}. Strings may thus promote pathogenesis in UPEC or other infections by enabling colonies to extend through polymer-laden environments such as mucus \cite{Witten2018mucuspermeability,Sauvaitre2022mucusbacteria}, wound matrices \cite{Percival2012biofilmsinwounds,Rahman2022wound}, or the biofilm matrix surrounding UPEC communities \cite{Kulkarni2026upecbiofilm}. Strings may also promote access to nutrients and oxygen with no interior region prone to metabolic limitation \cite{Wessel2014oxygen}. Growth into extended strings may help bacteria defend against phagocytosis from macrophages, since engulfment is slower for large \cite{Pratten1986size} and high-aspect-ratio \cite{Champion2006shape} targets. Second, in a string, each bacterium contacts only its two neighbors. The string structure may therefore limit the spread of any process that spreads by contact, including infection via phage \cite{Eriksen2020phage}, even though strings present more surface area per cell than a compact colony in which each cell has many neighbors. The bacteria in strings may be more exposed to host immune response \cite{Monack2004immune} or antimicrobials. Conversely, strings may promote processes that depend on cell-to-cell contact, including conjugation \cite{Babic2011conjugation}, contact-dependent growth inhibition \cite{Dessartine2024contact}, and short-range signaling \cite{Domenech2018shortrange,vanGestel2021shortrange}. 

Finally, the ability to tailor the morphology of bacterial networks by combining bacterial surface expression with physical interactions offers a route to grow spatially structured bacterial communities \cite{Paula2020spatialstructure}. Thus, our study suggests a route to engineer living materials \cite{Jin2020}  by integrating biological and synthetic interactions.

\section{\label{section:methods} Materials and Methods}

\subsection{\label{methods:strains} Strains, plasmids, and growth conditions}
Bacterial strains and plasmids used in this study are listed in Table SI in the Supplemental Material. We grow cells on Luria Bertani (LB) agar plates (containing 15 g L$^{-1}$  agar (BD Chemicals), 5 g L$^{-1}$  yeast extract (BD Chemicals), 10 g L$^{-1}$ tryptone (BD Chemicals), and 5 g L$^{-1}$  NaCl (VWR Chemicals)) or in LB broth (containing 5 g L$^{-1}$ yeast extract, 10 g L$^{-1}$  tryptone, and 5 g L$^{-1}$  NaCl) supplemented with 50 $\mu$g mL$^{-1}$ apramycin sulfate (Indofine Chemical Company) unless otherwise stated. 

\subsection{\label{methods:fliC knockout mutant}Construction of \textit{fliC} knockout mutant}

We construct the MG1655$\Delta$\textit{fliC} strain by deleting the \textit{fliC} gene from \textit{E.\ coli} K-12 MG1655 using the $\lambda$ Red recombinase-based method \cite{Datsenko2000}. Briefly, we design primers fliC-P1-FW and fliC-P1-RV to amplify the FRT-flanked kanamycin resistance gene from the template plasmid pKD13. We amplify the PCR product with a second set of primers (fliC-P2-FW and fliC-P2-RV) that contain 5' overhangs homologous to sequences flanking \textit{fliC} gene in the \textit{E. coli} MG1655  chromosome. We then purify and transform this second PCR product into an \textit{E. coli} MG1655 strain carrying the $\lambda$ Red helper plasmid pKD46. Following selection of kanamycin-resistant clones and confirmation of \textit{fliC} deletion by colony PCR, we remove the kanamycin resistance gene using helper plasmid pCP20. We finally confirm the \textit{fliC} deletion by colony PCR using primers fliC-test-FW and fliC-test-RV. 

\subsection{\label{methods:egfp}Construction of fluorescent plasmids}

We insert the gene encoding enhanced green fluorescent protein (\textit{egfp}) under the control of a \textit{tet} promoter in the p(blank) plasmid to directly visualize the \textit{fliC} knockout mutant carrying the p(blank) plasmid using confocal microscopy. We amplify the \textit{tet} promoter from the pPCC1234 plasmid \cite{Tang2008} using primers p1234-Ptet-XmnI-FW and p1234-Ptet-egfp-RV. We amplify the \textit{egfp} gene from the pPCC1234 plasmid using primers p1234-Ptet-egfp-FW and p1234-egfp-BsiWI-RV. We then linearly assemble and amplify the \textit{tet} promoter (P\textit{tet}) and \textit{egfp} gene using an overlap extension PCR with the end primers p1234-Ptet-XmnI-FW and p1234-egfp-BsiWI-RV. We finally ligate the P\textit{tet}-\textit{egfp} fragment into p(blank) digested with XmnI and BsiWI using T4 DNA ligase.

\vspace{\baselineskip}

We insert the \textit{mCherry} gene under the control of a \textit{tet} promoter in the p(\textit{fim}) plasmid that we previously constructed \cite{McLay2018} to directly visualize the fimbriated strain using confocal microscopy. We amplify the \textit{tet} promoter from the pPCC1234 plasmid using primers p1234-Ptet-XmnI-FW and p1234-Ptet-rfp-RV. We amplify the \textit{mCherry} gene from the pPCC2107 plasmid \cite{BahramiMoghadam2024} using primers p2107-Ptet-rfp-FW and p2107-rfp-BsiWI-RV. We linearly assemble and amplify the \textit{tet} promoter (P\textit{tet}) and \textit{mCherry} gene using an overlap extension PCR with the end primers p1234-Ptet-XmnI-FW and p2107-rfp-BsiWI-RV. We then ligate the P\textit{tet}-\textit{mCherry} fragment into p(\textit{fim}) digested with XmnI and BsiWI using T4 DNA ligase. The primers used in the construction of both fluorescent plasmids are provided in Table SII in the Supplemental Material.

\subsection{\label{methods:AFM}Atomic force microscopy (AFM)}
We use AFM to visually confirm the presence or absence of fimbriae. We attach freshly cleaved mica sheets (Electron Microscopy Sciences) to AFM specimen discs using non-conductive double-sided adhesive tabs (Microscopy Solutions) and coat them with 10 $\mu$L of 0.01\% poly-\textsc{L}-lysine (Sigma Aldrich). We grow overnight cultures from a single colony and subculture them the following morning, starting with an optical density at 600 nm (OD$_\mathrm{600}$) of 0.1. We induce the subcultures with desired concentration of isopropyl $\beta$-\textsc{D}-1-thiogalactopyranoside (IPTG) (Fisher Bioreagents) at an OD$_\mathrm{600}$ of 0.5. After 4 hours of induction, we centrifuge 1 mL of bacterial culture at 800 g for 10 min, discard the supernatant, and resuspend the cells in sterile MilliQ water. We then dilute the resuspended cells to an OD$_\mathrm{600}$ of 0.1 and add them to the poly-L-lysine coated mica sheets glued on specimen discs. We dry the samples for 1 hour and then scan them using a Bruker Multi-mode AFM equipped with an ScanAsyst Air tip (Bruker) in ScanAsyst Air mode. 
 
\subsection{\label{methods:polymer}Polymer solutions}

We prepare stock solutions of sodium polystyrene sulfonate (NaPSS) ($M_w = 70$ kDa, Sigma Aldrich) and Ficoll ($M_w = 400$ kDa, Research Products International) at concentrations of 20 g L$^{-1}$ and 100 g L$^{-1}$, respectively, in Phosphate Buffered Saline (PBS) for non-growth studies or in LB medium for growth studies. We filter the polymer stock solutions through a 0.2 $\mu$m disposable syringe filter prior to use and dilute them to desired concentrations. We determine the radius of gyration of the polymers at infinite dilution using intrinsic viscosity measurements (Supplemental Material V).

\subsection{\label{methods:sedimentation}Sedimentation assay}

We grow overnight cultures at 37$^\circ$C and 250 rpm from a single colony. We inoculate cells from the overnight cultures into fresh LB broth supplemented with 50 $\mu$g mL$^{-1}$ apramycin sulfate at 1:100 dilution and induce them with 10 $\mu$M IPTG at an OD$_\mathrm{600}$ of 0.5. After 4 h of induced expression, we wash the cells and resuspend them in PBS. We then mix the cell suspension and polymer solution at different concentrations, add 1 mL of the mixture to 1.5 mL disposable cuvettes (VWR) and homogenize them by thorough mixing. We visually score aggregation after 4 h at room temperature.

\subsection{\label{methods:imaging_nongrowth} Imaging bacterial aggregation in polymer solutions}
We grow overnight cultures at 37$^\circ$C and 250 rpm from a single colony. We inoculate cells from the overnight cultures into fresh LB broth supplemented with 50 $\mu$g mL$^{-1}$ apramycin sulfate at 1:100 dilution and induce them with 10 $\mu$M IPTG at an OD$_\mathrm{600}$ of 0.5. After 4 h of induced expression, we wash the cells in PBS and dilute them to an OD$_\mathrm{600}$ of 0.1 in the polymer solution at the concentration to be tested. We then add the samples to the base (20 mm diameter, $\sim 1$ mm height) of a transparent-walled glass-bottom Petri dish (35 mm in diameter and 10 mm in height, Cellvis) and image them using a Leica DMi8 inverted confocal microscope equipped with a 63x oil immersion lens (NA = 1.4). 

\subsection{\label{methods:imaging_growth} Imaging bacterial proliferation in polymer solutions}
We grow overnight cultures at 37$^\circ$C and 250 rpm from a single colony. We inoculate cells from the overnight cultures into fresh LB broth supplemented with 50 $\mu$g mL$^{-1}$ apramycin sulfate at 1:100 dilution and induce them with 0, 10, or 50 $\mu$M IPTG at an OD$_\mathrm{600}$ of 0.5. After 4 h of induction, we dilute the cultures to a concentration of $\approx 4 \times 10^5$ cells mL$^{-1}$ in the polymer solution, supplemented with 50 $\mu$g mL$^{-1}$ apramycin sulfate and the indicated concentration of IPTG, at the concentration to be tested and add them to the base of a transparent-walled glass-bottom Petri dish. We then seal the Petri dish with an overlying circular polydimethylsiloxane (PDMS) slab to minimize evaporation while still allowing oxygen to be available to the proliferating cells. We image the sealed chamber from below using a Leica DMi8 inverted confocal microscope equipped with a 63x oil immersion lens (N.A.\ 1.4) and maintained at 37$^\circ$C for 5 to 6 h. We obtain 3-D $z$-stacks by sequentially acquiring 2-D images from the bottom of the Petri dish up to a maximum height of 50 $\mu$m separated by $\Delta z$ = 0.5 $\mu$m.  

\subsection{\label{methods:flow_cytometry} Flow cytometry}

We grow overnight cultures at 37$^\circ$C and 250 rpm from a single colony. We inoculate cells from the overnight cultures into fresh LB broth supplemented with 50 $\mu$g mL$^{-1}$ apramycin sulfate at 1:100 dilution and induce them with 0, 10, or 50 $\mu$M IPTG at an OD$_\mathrm{600}$ of 0.5. After 4 h of induced expression, we wash the cells in PBS and dilute them to a cell density of $\sim10^{6}$ cells mL$^{-1}$ in flow cytometry tubes (5 mL round-bottom Falcon tubes (BD Biosciences), size: $12 \times 75$ mm). We treat the resulting cell suspensions with 20 mM propidium iodide (PI) (Thermo Fisher Scientific) and 1 mM Redox Sensor Green (RSG) (Thermo Fisher Scientific) dyes and incubate them in the dark at 37$^\circ$C for 15 min. We then analyze the samples using a conventional bench-top flow cytometer (NovoCyte 3000RYB, ACEA Biosciences Inc.) with a sample flow rate of 14 $\mu$L min$^{-1}$ and a sheath flow rate of 6.5 $\mu$L min$^{-1}$. We excite the PI and RSG with 561 nm and 488 nm wavelength lasers, respectively and detect the emitted fluorescence with 615/20 nm and 530/30 nm bandpass filters, respectively. We record at least 30,000 events for each sample. We collect the data using NovoExpress software. Dead cells, obtained by treating live cells with 70 \% v/v ethanol, are used as a positive control.

\subsection{\label{methods:agent} Agent-based simulations}

We perform two-dimensional agent-based Brownian dynamics simulations of proliferating bacteria, extending the model developed previously to describe 
cable formation of non-fimbriated cells~\cite{GonzalezLaCorte2025} with the addition of fimbria-induced steric repulsion. Details are provided in the Supplemental Materials (IX). In brief, each cell is modeled as a soft spherocylinder that elongates exponentially at fixed diameter and divides into two daughter cells of equal length after a doubling time $t_d$. Adjacent cells interact through a Hertzian elastic repulsion that prevents their bodies from overlapping, a polymer-induced depletion attraction whose magnitude depends on the polymer size and on the sizes and relative orientations of the two cells~\cite{Asakura1958,savenko2006phase}, and a soft repulsion generated when their fimbriated exclusion regions overlap. Each cell additionally undergoes translational and rotational Brownian motion, with mobilities set by the cell length and orientation and the solution viscosity.       

\subsection{\label{methods:classify}Classification of proliferation states}

We classify the states formed by bacteria as they proliferate in confocal micrographs or agent-based simulations (both of dimensions 71 $\mu$m $\times$ 71 $\mu$m, corresponding to 35 cell lengths in $x$ and $y$) using four heuristics. (a) In `dispersed' states, the great majority of cells remain individually dispersed, and aggregates contain fewer than 6 bacteria. (b) `Clusters' contain more than more than six bacteria (with side-by-side or end-to-end attractions) but that did not span the field of view. (c) In `cable' states, cells form space-spanning structures across the field of view and were connected side-by-side and end-to-end. (d) In `string' states, cells form space-spanning structures across the field of view but were only connected end-to-end.

\section{Data Availability}

The experimental and simulation data for this study are available in a Zenodo data repository (doi: 10.5281/zenodo.24592135). The agent-based model and its implementation are specified in full in Supplementary Material IX.

\begin{acknowledgments}
% \SSD{We thank Sebastian Gonzalez La Corte for providing the simulation code on which the present model is based, and for useful discussions at the inception of this work.} 
We thank the National Science Foundation (DMR-2104796, to JCC and PCC, and CBET-1941716 and EF-2124863 to SSD), the Welch Foundation (E-1869-2024-0404, to JCC), and the Camille Dreyfus Teacher-Scholar and Pew Biomedical Scholars Programs (to SSD) for support. SSD used Claude (Anthropic, Opus 5) to assist with literature searching and for revising the introduction. The authors reviewed and verified all content and take
full responsibility for the final manuscript.
\end{acknowledgments}

\bibliography{bacteria}% Produces the bibliography via BibTeX.

@article{10.1021/acs.biochem.9b01101,
    author = {Chu, I-Te and Speer, Shannon L. and Pielak, Gary J.},
    title = {Rheostatic Control of Protein Expression Using Tuner
Cells},
    journal = {Biochemistry},
    volume = {59},
    number = {6},
    pages = {733-735},
    year = {2020},
    month = {02},
    issn = {0006-2960},
    doi = {10.1021/acs.biochem.9b01101}
}

@article{BHOGALE2022183,
title = {Switching off: The phenotypic transition to the uninduced state of the lactose uptake pathway},
journal = {Biophysical Journal},
volume = {121},
number = {2},
pages = {183-192},
year = {2022},
issn = {0006-3495},
doi = {https://doi.org/10.1016/j.bpj.2021.12.027},
url = {https://www.sciencedirect.com/science/article/pii/S0006349521039515},
author = {Prasanna M. Bhogale and Robin A. Sorg and Jan-Willem Veening and Johannes Berg}
}

@article{flemming2010biofilm,
  title={The biofilm matrix},
  author={Flemming, Hans-Curt and Wingender, Jost},
  journal={Nature reviews microbiology},
  volume={8},
  number={9},
  pages={623--633},
  year={2010},
  publisher={Nature Publishing Group UK London}
}

@article{anderson2003intracellular,
  title={Intracellular bacterial biofilm-like pods in urinary tract infections},
  author={Anderson, Gregory G and Palermo, Joseph J and Schilling, Joel D and Roth, Robyn and Heuser, John and Hultgren, Scott J},
  journal={Science},
  volume={301},
  number={5629},
  pages={105--107},
  year={2003},
  publisher={American Association for the Advancement of Science}
}

@article{Cheah1987,
   author = {U. Eong Cheah and William A. Weigand and Benjamin C. Stark},
   doi = {10.1016/0147-619X(87)90040-0},
   issn = {0147619X},
   issue = {2},
   journal = {Plasmid},
   month = {9},
   pages = {127-134},
   title = {Effects of recombinant plasmid size on cellular processes in \textit{Escherichia coli}},
   volume = {18},
   year = {1987}
}

@article{Tsang2017,
   author = {Jennifer Tsang},
   issn = {2328-9791},
   issue = {5},
   journal = {Postdoc Journal},
   month = {5},
   pages = {3-9},
   pmid = {28781980},
   title = {Bacterial plasmid addiction systems and their implications for antibiotic drug development.},
   volume = {5},
   year = {2017}
}

@article{BahramiMoghadam2024,
   author = {Ehsan Bahrami Moghadam and Nam Nguyen and Yixi Wang and Patrick C. Cirino},
   doi = {10.3390/bios14070325},
   issn = {2079-6374},
   issue = {7},
   journal = {Biosensors},
   month = {6},
   pages = {325},
   title = {Directed Evolution of Protein-Based Sensors for Anaerobic Biological Activation of Methane},
   volume = {14},
   year = {2024}
}

@article{Wurpel2013,
   author = {Daniël J. Wurpel and Scott A. Beatson and Makrina Totsika and Nicola K. Petty and Mark A. Schembri},
   doi = {10.1371/journal.pone.0052835},
   issn = {1932-6203},
   issue = {1},
   journal = {PLoS ONE},
   month = {1},
   pages = {e52835},
   title = {Chaperone-Usher Fimbriae of \textit{Escherichia coli}},
   volume = {8},
   year = {2013}
}

@article{Kalyuzhnaya2008,
   author = {Marina G Kalyuzhnaya and Mary E Lidstrom and Ludmila Chistoserdova},
   doi = {10.1038/ismej.2008.32},
   issn = {1751-7362},
   issue = {7},
   journal = {The ISME Journal},
   month = {7},
   pages = {696-706},
   title = {Real-time detection of actively metabolizing microbes by redox sensing as applied to methylotroph populations in {Lake Washington}},
   volume = {2},
   year = {2008}
}

@article{savenko2006phase,
  title={Phase behavior of a suspension of colloidal hard rods and nonadsorbing polymer},
  author={Savenko, S. V. and Dijkstra, Marjolein},
  journal={Journal of Chemical Physics},
  volume={124},
  number={23},
  year={2006},
  publisher={AIP Publishing}, 
  pages = {234902}
}

@article{verma1998entropic,
  title={Entropic colloidal interactions in concentrated {DNA} solutions},
  author={Verma, Ritu and Crocker, John C and Lubensky, Tom C and Yodh, Arjun G},
  journal={Physical Review Letters},
  volume={81},
  number={18},
  pages={4004},
  year={1998},
  publisher={APS}
}

@article{Kragh2016,
author = {Kasper N. Kragh and Jaime B. Hutchison and Gavin Melaugh and Chris Rodesney and Aled E. L. Roberts and Yasuhiko Irie and Peter {\O}. Jensen and Stephen P. Diggle and Rosalind J. Allen and Vernita Gordon and Thomas Bjarnsholt},
title = {Role of Multicellular Aggregates in Biofilm Formation},
journal = {mBio},
volume = {7},
number = {2},
pages = {00237},
year = {2016},
doi = {10.1128/mbio.00237-16}
}

@article{Simon-Soro2022,
   author = {Aurea Simon-Soro and Zhi Ren and Bastiaan P. Krom and Michel A. Hoogenkamp and Pedro J. Cabello-Yeves and Scott G. Daniel and Kyle Bittinger and Inmaculada Tomas and Hyun Koo and Alex Mira},
   doi = {10.1128/mbio.00131-22},
   issn = {2150-7511},
   issue = {1},
   journal = {mBio},
   month = {2},
   title = {Polymicrobial Aggregates in Human Saliva Build the Oral Biofilm},
   volume = {13},
   year = {2022},
   pages = {e00131-22}
}

@article{Melaugh2016,
   author = {Gavin Melaugh and Jaime Hutchison and Kasper Nørskov Kragh and Yasuhiko Irie and Aled Roberts and Thomas Bjarnsholt and Stephen P. Diggle and Vernita D. Gordon and Rosalind J. Allen},
   doi = {10.1371/journal.pone.0149683},
   issn = {1932-6203},
   issue = {3},
   journal = {PLOS ONE},
   month = {3},
   pages = {e0149683},
   title = {Shaping the Growth Behaviour of Biofilms Initiated from Bacterial Aggregates},
   volume = {11},
   year = {2016}
}

@article{Kragh2023,
   author = {Kasper N. Kragh and Tim Tolker-Nielsen and Mads Lichtenberg},
   doi = {10.1038/s42003-023-05281-4},
   issn = {2399-3642},
   issue = {1},
   journal = {Communications Biology},
   month = {9},
   pages = {898},
   title = {The non-attached biofilm aggregate},
   volume = {6},
   year = {2023}
}

@article{MarkWelch2016,
   author = {Jessica L. Mark Welch and Blair J. Rossetti and Christopher W. Rieken and Floyd E. Dewhirst and Gary G. Borisy},
   issn = {0027-8424},
   issue = {6},
   journal = {Proceedings of the National Academy of Sciences},
   month = {2},
   title = {Biogeography of a human oral microbiome at the micron scale},
   volume = {113},
   pages = {E791},
   year = {2016}
}

@article{Puri2024,
   author = {Devina Puri and Kyle R. Allison},
   issn = {0027-8424},
   issue = {23},
   journal = {Proceedings of the National Academy of Sciences},
   month = {6},
   title = {\textit{Escherichia coli} self-organizes developmental rosettes},
   volume = {121},
   pages = {e2315850121},
   year = {2024}
}

@article{Brinton1965,
   author = {Charles C. Brinton},
   doi = {10.1111/j.2164-0947.1965.tb02342.x},
   issn = {0028-7113},
   issue = {8 Series II},
   journal = {Transactions of the New York Academy of Sciences},
   month = {6},
   pages = {1003-1054},
   title = {THE STRUCTURE, FUNCTION, SYNTHESIS AND GENETIC CONTROL OF BACTERIAL PILI AND A MOLECULAR MODEL FOR {DNA} AND {RNA} TRANSPORT IN {Gram} negative BACTERIA},
   volume = {27},
   year = {1965}
}

@article{Hospenthal2017,
   author = {Manuela K. Hospenthal and Dawid Zyla and Tiago R.D. Costa and Adam Redzej and Christoph Giese and James Lillington and Rudi Glockshuber and Gabriel Waksman},
   doi = {10.1016/j.str.2017.10.004},
   issn = {09692126},
   issue = {12},
   journal = {Structure},
   month = {12},
   pages = {1829-1838.e4},
   title = {The Cryoelectron Microscopy Structure of the Type 1 Chaperone-Usher Pilus Rod},
   volume = {25},
   year = {2017}
}

@article{berne2015adhesins,
author = {Cécile Berne and Adrien Ducret and Gail G. Hardy and Yves V. Brun},
title = {Adhesins Involved in Attachment to Abiotic Surfaces by {Gram}-Negative Bacteria},
journal = {Microbiology Spectrum},
volume = {3},
number = {4},
pages = {10.1128/microbiolspec.mb-0018-2015},
year = {2015},
doi = {10.1128/microbiolspec.mb-0018-2015}
}

@article{forero2006uncoiling,
  title={Uncoiling mechanics of \textit{Escherichia coli} {type I} fimbriae are optimized for catch bonds},
  author={Forero, Manu and Yakovenko, Olga and Sokurenko, Evgeni V and Thomas, Wendy E and Vogel, Viola},
  journal={PLoS Biology},
  volume={4},
  number={9},
  pages={e298},
  year={2006},
  publisher={Public Library of Science, San Francisco, USA}
}

@article{Dorken2012,
   author = {Gary Dorken and Gail P. Ferguson and Chris E. French and Wilson C. K. Poon},
   doi = {10.1098/rsif.2012.0498},
   issn = {1742-5689},
   issue = {77},
   journal = {Journal of The Royal Society Interface},
   month = {12},
   pages = {3490-3502},
   title = {Aggregation by depletion attraction in cultures of bacteria producing exopolysaccharide},
   volume = {9},
   year = {2012}
}

@article{Schwarz-Linek2010SM,
   author = {Jana Schwarz-Linek and Alexander Winkler and Laurence G. Wilson and Nhan T. Pham and Tanja Schilling and Wilson C. K. Poon},
   doi = {10.1039/c0sm00214c},
   issn = {1744-683X},
   issue = {18},
   journal = {Soft Matter},
   pages = {4540},
   title = {Polymer-induced phase separation in \textit{Escherichia coli} suspensions},
   volume = {6},
   year = {2010}
}

@article{Schwarz-Linek2010EPL,
   author = {J. Schwarz-Linek and G. Dorken and A. Winkler and L. G. Wilson and N. T. Pham and C. E. French and T. Schilling and W. C. K. Poon},
   doi = {10.1209/0295-5075/89/68003},
   issn = {0295-5075},
   issue = {6},
   journal = {EPL (Europhysics Letters)},
   month = {3},
   pages = {68003},
   title = {Polymer-induced phase separation in suspensions of bacteria},
   volume = {89},
   year = {2010}
}

@article{RameshKumar2024,
   author = {Udayanidhi {Ramesh Kumar} and Nam T. Nguyen and Narendra K. Dewangan and Sayed Golam Mohiuddin and Mehmet A. Orman and Patrick C. Cirino and Jacinta C. Conrad},
   doi = {10.1039/d4sm00499j},
   issn = {17446848},
   issue = {37},
   journal = {Soft Matter},
   month = {7},
   pages = {7397-7404},
   pmid = {39021099},
   publisher = {Royal Society of Chemistry},
   title = {Co-Expression of type 1 fimbriae and flagella in \textit{Escherichia coli}: consequences for adhesion at interfaces},
   volume = {20},
   year = {2024}
}

@article{Tang2008,
   author = {Shuang-Yan Tang and Hossein Fazelinia and Patrick C. Cirino},
   doi = {10.1021/ja7109053},
   issn = {0002-7863},
   issue = {15},
   journal = {Journal of the American Chemical Society},
   month = {4},
   pages = {5267-5271},
   title = {{AraC} Regulatory Protein Mutants with Altered Effector Specificity},
   volume = {130},
   year = {2008}
}

@article{Secor2018,
   author = {Patrick R. Secor and Lia A. Michaels and Anina Ratjen and Laura K. Jennings and Pradeep K. Singh},
   doi = {10.1073/pnas.1806005115},
   issn = {0027-8424},
   issue = {42},
   journal = {Proceedings of the National Academy of Sciences},
   month = {10},
   pages = {10780-10785},
   title = {Entropically driven aggregation of bacteria by host polymers promotes antibiotic tolerance in \textit{Pseudomonas aeruginosa}},
   volume = {115},
   year = {2018}
}

@article{Jennings2021,
   author = {Laura K. Jennings and Julia E. Dreifus and Courtney Reichhardt and Kelly M. Storek and Patrick R. Secor and Daniel J. Wozniak and Katherine B. Hisert and Matthew R. Parsek},
   doi = {10.1016/j.celrep.2021.108782},
   issn = {22111247},
   issue = {8},
   journal = {Cell Reports},
   month = {2},
   pages = {108782},
   title = {\textit{Pseudomonas aeruginosa} aggregates in cystic fibrosis sputum produce exopolysaccharides that likely impede current therapies},
   volume = {34},
   year = {2021}
}

@article{GonzalezLaCorte2025,
   author = {Sebastian {Gonzalez La Corte} and Corey A. Stevens and Gerardo Cárcamo-Oyarce and Katharina Ribbeck and Ned S. Wingreen and Sujit S. Datta},
   doi = {10.1126/sciadv.adq7797},
   issn = {2375-2548},
   issue = {3},
   journal = {Science Advances},
   month = {1},
   title = {Morphogenesis of bacterial cables in polymeric environments},
   pages = {eadq7797},
   volume = {11},
   year = {2025}
}

@article{Schwarz-Linek2012,
   author = {J. Schwarz-Linek and C. Valeriani and A. Cacciuto and M. E. Cates and D. Marenduzzo and A. N. Morozov and W. C. K. Poon},
   doi = {10.1073/pnas.1116334109},
   issn = {0027-8424},
   issue = {11},
   journal = {Proceedings of the National Academy of Sciences},
   month = {3},
   pages = {4052-4057},
   title = {Phase separation and rotor self-assembly in active particle suspensions},
   volume = {109},
   year = {2012}
}

@article{Eboigbodin2005,
   author = {K. E. Eboigbodin and J. R. A. Newton and A. F. Routh and C. A. Biggs},
   doi = {10.1021/la051740u},
   issn = {0743-7463},
   issue = {26},
   journal = {Langmuir},
   month = {12},
   pages = {12315-12319},
   title = {Role of Nonadsorbing Polymers in Bacterial Aggregation},
   volume = {21},
   year = {2005}
}

@article{Secor2022,
   author = {Patrick R. Secor and Lia A. Michaels and DeAnna C. Bublitz and Laura K. Jennings and Pradeep K. Singh},
   doi = {10.3389/fcimb.2022.869736},
   issn = {2235-2988},
   journal = {Frontiers in Cellular and Infection Microbiology},
   month = {6},
   title = {The Depletion Mechanism Actuates Bacterial Aggregation by Exopolysaccharides and Determines Species Distribution \& Composition in Bacterial Aggregates},
   volume = {12},
   pages = {869736},
   year = {2022}
}

@article{Peters2021,
   author = {V. F. D. Peters and M. Vis and R. Tuinier and H. N. W. Lekkerkerker},
   doi = {10.1063/5.0045435},
   issn = {0021-9606},
   issue = {15},
   journal = {Journal of Chemical Physics},
   month = {4},
   title = {Phase separation in mixed suspensions of bacteria and nonadsorbing polymers},
   volume = {154},
   year = {2021},
   pages = {151101}
}

@article{Asakura1958,
   author = {Sho Asakura and Fumio Oosawa},
   doi = {10.1002/pol.1958.1203312618},
   issn = {0022-3832},
   issue = {126},
   journal = {Journal of Polymer Science},
   month = {12},
   pages = {183-192},
   title = {Interaction between particles suspended in solutions of macromolecules},
   volume = {33},
   year = {1958}
}

@article{Vasconcellos1994,
   author = {Carol A. Vasconcellos and Philip G. Allen and Mary Ellen Wohl and Jeffrey M. Drazen and Paul A. Janmey and Thomas P. Stossel},
   doi = {10.1126/science.8310295},
   issn = {0036-8075},
   issue = {5149},
   journal = {Science},
   month = {2},
   pages = {969-971},
   title = {Reduction in Viscosity of Cystic Fibrosis Sputum in Vitro by Gelsolin},
   volume = {263},
   year = {1994}
}

@article{Brandt1995,
   author = {T Brandt and S Breitenstein and H von der Hardt and B Tummler},
   doi = {10.1136/thx.50.8.880},
   issn = {0040-6376},
   issue = {8},
   journal = {Thorax},
   month = {8},
   pages = {880-882},
   title = {{DNA} concentration and length in sputum of patients with cystic fibrosis during inhalation with recombinant human {DNase}},
   volume = {50},
   year = {1995}
}

@article{Fahy2010,
   author = {John V. Fahy and Burton F. Dickey},
   doi = {10.1056/NEJMra0910061},
   issn = {0028-4793},
   issue = {23},
   journal = {New England Journal of Medicine},
   month = {12},
   pages = {2233-2247},
   title = {Airway Mucus Function and Dysfunction},
   volume = {363},
   year = {2010}
}

@book{Clark1988,
   city = {Boston, MA},
   doi = {10.1007/978-1-4899-0185-9},
   editor = {Richard A. F. Clark},
   isbn = {978-1-4899-0187-3},
   publisher = {Springer US},
   title = {The Molecular and Cellular Biology of Wound Repair},
   year = {1988}
}

@article{Decho2017,
   author = {Alan W. Decho and Tony Gutierrez},
   doi = {10.3389/fmicb.2017.00922},
   issn = {1664-302X},
   journal = {Frontiers in Microbiology},
   month = {5},
   title = {Microbial Extracellular Polymeric Substances {(EPSs)} in Ocean Systems},
   volume = {8},
   year = {2017},
   pages = {922}
}

@article{McShane2021,
   author = {Abigail McShane and Jade Bath and Ana M. Jaramillo and Caroline Ridley and Agnes A. Walsh and Christopher M. Evans and David J. Thornton and Katharina Ribbeck},
   doi = {10.1016/j.cub.2021.06.093},
   issn = {09609822},
   issue = {15},
   journal = {Current Biology},
   month = {8},
   pages = {R938-R945},
   title = {Mucus},
   volume = {31},
   year = {2021}
}

@article{Datsenko2000,
   author = {Kirill A. Datsenko and Barry L. Wanner},
   doi = {10.1073/pnas.120163297},
   issn = {0027-8424},
   issue = {12},
   journal = {Proceedings of the National Academy of Sciences},
   month = {6},
   pages = {6640-6645},
   title = {One-step inactivation of chromosomal genes in \textit{Escherichia coli} {{K-12}} using PCR products},
   volume = {97},
   year = {2000}
}

@article{Boulos1999,
   author = {Lina Boulos and Michèle Prévost and Benoit Barbeau and Josée Coallier and Raymond Desjardins},
   doi = {10.1016/S0167-7012(99)00048-2},
   issn = {01677012},
   issue = {1},
   journal = {Journal of Microbiological Methods},
   month = {7},
   pages = {77-86},
   title = {{LIVE/DEAD®} {BacLight™}: application of a new rapid staining method for direct enumeration of viable and total bacteria in drinking water},
   volume = {37},
   year = {1999}
}

@article{Qiao2021,
author = {Jun Qiao and Xin Tan and Hongyu Ren and Zheng Wu and Xiaoqing Hu and Xiaoyuan Wang},
year = 2021,
title = {Construction of an \textit{Escherichia coli} Strain Lacking Fimbriae by Deleting 64 Genes and Its Application for Efficient Production of Poly(3-Hydroxybutyrate) and \textsc{L}-Threonine},
journal = {Applied and Environmental Microbiology},
volume = {87},
pages = {e00381-21}
}

@article{Jin2020,
    doi = {10.1371/journal.pone.0243280},
    author = {Jin, Xing AND Marshall, Jeffrey S.},
    journal = {PLOS ONE},
    publisher = {Public Library of Science},
    title = {Mechanics of biofilms formed of bacteria with fimbriae appendages},
    year = {2020},
    month = {12},
    volume = {15},
    url = {https://doi.org/10.1371/journal.pone.0243280},
    pages = {e0243280},
    number = {12},
}

@article{connell1996type,
  title={Type 1 fimbrial expression enhances \textit{Escherichia coli} virulence for the urinary tract},
  author={Connell, I and Agace, William and Klemm, Per and Schembri, Mark and M{\u{a}}rild, S and Svanborg, Catharina},
  journal={Proceedings of the National Academy of Sciences},
  volume={93},
  number={18},
  pages={9827--9832},
  year={1996}
}

@article{Hancock2011,
   author = {Viktoria Hancock and Ingun Lund Witsø and Per Klemm},
   doi = {10.1016/j.ijmm.2011.04.018},
   issn = {14384221},
   issue = {7},
   journal = {International Journal of Medical Microbiology},
   month = {11},
   pages = {570-576},
   pmid = {21646046},
   title = {Biofilm formation as a function of adhesin, growth medium, substratum and strain type},
   volume = {301},
   year = {2011}
}

@article{Pratt1998,
   author = {Leslie A. Pratt and Roberto Kolter},
   doi = {10.1046/j.1365-2958.1998.01061.x},
   issn = {0950382X},
   issue = {2},
   journal = {Molecular Microbiology},
   pages = {285-293},
   pmid = {9791174},
   title = {Genetic analysis of \textit{Escherichia coli} biofilm formation: Roles of flagella, motility, chemotaxis and type {I} pili},
   volume = {30},
   year = {1998}
}

@article{Rodrigues2009,
   author = {Debora F. Rodrigues and Menachem Elimelech},
   doi = {10.1080/08927010902833443},
   issn = {08927014},
   issue = {5},
   journal = {Biofouling},
   month = {7},
   pages = {401-411},
   pmid = {19306144},
   title = {Role of type 1 fimbriae and mannose in the development of \textit{Escherichia coli} K12 biofilm: From initial cell adhesion to biofilm formation},
   volume = {25},
   year = {2009}
}

@article{AvalosVizcarra2016,
	author = {Avalos Vizcarra, Ima and Hosseini, Vahid and Kollmannsberger, Philip and Meier, Stefanie and Weber, Stefan S. and Arnoldini, Markus and Ackermann, Martin and Vogel, Viola},
	journal = {Scientific Reports},
	number = {1},
	pages = {18109},
	title = {How type 1 fimbriae help \textit{Escherichia coli} to evade extracellular antibiotics},
	volume = {6},
	year = {2016}}

@article{McLay2018,
   author = {Ryan B. McLay and Hang N. Nguyen and Yuly Andrea Jaimes-Lizcano and Narendra K. Dewangan and Simone Alexandrova and Debora F. Rodrigues and Patrick C. Cirino and Jacinta C. Conrad},
   doi = {10.1021/acs.langmuir.7b02447},
   issn = {15205827},
   issue = {3},
   journal = {Langmuir},
   month = {1},
   pages = {1133-1142},
   pmid = {28976770},
   publisher = {American Chemical Society},
   title = {Level of Fimbriation Alters the Adhesion of \textit{Escherichia coli} Bacteria to Interfaces},
   volume = {34},
   year = {2018}
}

@article{Staerk2015expression,
    author = {Stærk, Kristian and Khandige, Surabhi and Kolmos, Hans Jørn and Møller-Jensen, Jakob and Andersen, Thomas Emil},
    title = {Uropathogenic \textit{Escherichia coli} Express Type 1 Fimbriae Only in Surface Adherent Populations Under Physiological Growth Conditions},
    journal = {Journal of Infectious Diseases},
    volume = {213},
    number = {3},
    pages = {386-394},
    year = {2015},
    month = {08},
    issn = {0022-1899},
    doi = {10.1093/infdis/jiv422}
}

@article{Eriksen2020phage,
	author = {Eriksen, Rasmus Skytte and Mitarai, Namiko and Sneppen, Kim},
	journal = {Biophysical Journal},
	month = {2026/08/10},
	number = {9},
	pages = {1896--1904},
	title = {On Phage Adsorption to Bacterial Chains},
	volume = {119},
	year = {2020}}

@article{vanGestel2021shortrange,
	author = {van Gestel, Jordi and Bareia, Tasneem and Tenennbaum, Bar and Dal Co, Alma and Guler, Polina and Aframian, Nitzan and Puyesky, Shani and Grinberg, Ilana and D'Souza, Glen G. and Erez, Zohar and Ackermann, Martin and Eldar, Avigdor},
	journal = {Nature Communications},
	number = {1},
	pages = {2324},
	title = {Short-range quorum sensing controls horizontal gene transfer at micron scale in bacterial communities},
	volume = {12},
	year = {2021}}

@article{Domenech2018shortrange,
	author = {Domenech, Arnau and Slager, Jelle and Veening, Jan-Willem},
	journal = {Cell Reports},
	month = {2026/08/10},
	number = {9},
	pages = {2390--2400.e3},
	title = {Antibiotic-Induced Cell Chaining Triggers Pneumococcal Competence by Reshaping Quorum Sensing to Autocrine-Like Signaling},
	volume = {25},
	year = {2018}}

@article{Babic2011conjugation,
author = {Ana Babic and Melanie B. Berkmen and Catherine A. Lee and Alan D. Grossman},
title = {Efficient Gene Transfer in Bacterial Cell Chains},
journal = {mBio},
volume = {2},
number = {2},
pages = {10.1128/mbio.00027-11},
year = {2011}
}

@article{Dessartine2024contact,
author = {Margot Marie Dessartine and Artemis Kosta and Thierry Doan and Éric Cascales and Jean-Philippe Côté},
title = {Type 1 fimbriae-mediated collective protection against type 6 secretion system attacks},
journal = {mBio},
volume = {15},
number = {4},
pages = {e02553-23},
year = {2024}
}

@article{Paula2020spatialstructure,
	author = {Paula, Amauri J. and Hwang, Geelsu and Koo, Hyun},
	journal = {Nature Communications},
	number = {1},
	pages = {1354},
	title = {Dynamics of bacterial population growth in biofilms resemble spatial and structural aspects of urbanization},
	volume = {11},
	year = {2020}}

@article{Higgs2025motility,
	author = {Matthew G. Higgs and Matthew A. Greenwald and Cristian Roca and Jade K. Macdonald and Ashelyn E. Sidders and Brian P. Conlon and Matthew C. Wolfgang},
	journal = {mBio},
	number = {6},
	title = {Flagellar motility and the mucus environment influence aggregation-mediated antibiotic tolerance of \textit{Pseudomonas aeruginosa} in chronic lung infection},
	volume = {16},
	year = {2025}}

@article{Chantratita2007morphology,
	author = {Narisara Chantratita and Vanaporn Wuthiekanun and Khaemaporn Boonbumrung and Rachaneeporn Tiyawisutsri and Mongkol Vesaratchavest and Direk Limmathurotsakul and Wirongrong Chierakul and Surasakdi Wongratanacheewin and Sasithorn Pukritiyakamee and Nicholas J. White and Nicholas P. J. Day and Sharon J. Peacock},
	journal = {Journal of Bacteriology},
	number = {3},
	pages = {807-817},
	title = {Biological Relevance of Colony Morphology and Phenotypic Switching by \textit{Burkholderia pseudomallei}},
	volume = {189},
	year = {2007}}

@article{you2018geometry,
  title={Geometry and mechanics of microdomains in growing bacterial colonies},
  author={You, Zhihong and Pearce, Daniel JG and Sengupta, Anupam and Giomi, Luca},
  journal={Physical Review X},
  volume={8},
  number={3},
  pages={031065},
  year={2018},
  publisher={APS}
}

@article{Witten2018mucuspermeability,
title = {Selective permeability of mucus barriers},
journal = {Current Opinion in Biotechnology},
volume = {52},
pages = {124-133},
year = {2018},
issn = {0958-1669},
doi = {https://doi.org/10.1016/j.copbio.2018.03.010},
url = {https://www.sciencedirect.com/science/article/pii/S095816691730232X},
author = {Jacob Witten and Tahoura Samad and Katharina Ribbeck}
}

@article{Rahman2022wound,
	author = {Rahman, Minhaz Ur and Fleming, Derek F. and Wang, Liyun and Rumbaugh, Kendra P. and Gordon, Vernita D. and Christopher, Gordon F.},
	journal = {npj Biofilms and Microbiomes},
	number = {1},
	pages = {49},
	title = {Microrheology of \textit{Pseudomonas aeruginosa} biofilms grown in wound beds},
	volume = {8},
	year = {2022}}

@article{Kulkarni2026upecbiofilm,
	author = {Kulkarni, Rutuparna and Lian, Zheng Jie and Nhu, Nguyen Thi Khanh and Blaskovich, Mark A. T. and Wohland, Thorsten and Schembri, Mark A.},
	journal = {npj Biofilms and Microbiomes},
	title = {The biofilm matrix of uropathogenic \textit{Escherichia coli} contributes to antimicrobial resistance by reducing molecular transport},
	year = {2026}}

@article{Sauvaitre2022mucusbacteria,
	author = {Sauvaitre, Thomas and Van Landuyt, Josefien and Durif, Claude and Roussel, Charl{\`e}ne and Sivignon, Adeline and Chalancon, Sandrine and Uriot, Oph{\'e}lie and Van Herreweghen, Florence and Van de Wiele, Tom and Etienne-Mesmin, Lucie and Blanquet-Diot, St{\'e}phanie},
	journal = {npj Biofilms and Microbiomes},
	number = {1},
	pages = {86},
	title = {Role of mucus-bacteria interactions in Enterotoxigenic \textit{Escherichia coli} {(ETEC) H10407} virulence and interplay with human microbiome},
	volume = {8},
	year = {2022}}

@article{Percival2012biofilmsinwounds,
author = {Percival, Steven L. and Hill, Katja E. and Williams, David W. and Hooper, Samuel J. and Thomas, Dave W. and Costerton, John W.},
title = {A review of the scientific evidence for biofilms in wounds},
journal = {Wound Repair and Regeneration},
volume = {20},
number = {5},
pages = {647-657},
year = {2012}
}

@article{Monack2004immune,
	author = {Monack, Denise M. and Mueller, Anne and Falkow, Stanley},
	journal = {Nature Reviews Microbiology},
	number = {9},
	pages = {747--765},
	title = {Persistent bacterial infections: the interface of the pathogen and the host immune system},
	volume = {2},
	year = {2004}}

@article{Wessel2014oxygen,
author = {Aimee K. Wessel and Talha A. Arshad and Mignon Fitzpatrick and Jodi L. Connell and Roger T. Bonnecaze and Jason B. Shear and Marvin Whiteley},
title = {Oxygen Limitation within a Bacterial Aggregate},
journal = {mBio},
volume = {5},
number = {2},
pages = {10.1128/mbio.00992-14},
year = {2014}
}

@book{kim2013microhydrodynamics,
  title={Microhydrodynamics: principles and selected applications},
  author={Kim, S and Karrila, SJ},
  publisher = {Dover Publications},
  year={2013}
}

@article{lobo2021polymer,
  title={Polymer-induced microcolony compaction in early biofilms: A computer simulation study},
  author={Lobo-Cabrera, Francisco Javier and Patti, Alessandro and Govantes, Fernando and Cuetos, Alejandro},
  journal={Physical Review E},
  volume={103},
  number={5},
  pages={052407},
  year={2021},
  publisher={APS}
}

@article{shimizu1962effect,
  title={Effect of molecular shape on nuclear magnetic relaxation},
  author={Shimizu, Hiroshi},
  journal={Journal of Chemical Physics},
  volume={37},
  number={4},
  pages={765--778},
  year={1962},
  publisher={American Institute of Physics}
}

@article{GonzalezLaCore2026chains,
author = {Sebastian {Gonzalez La Corte}  and Thomas G. J. Chandler  and Saverio E. Spagnolie  and Ned S. Wingreen  and Sujit S. Datta },
title = {Morphogenesis of bacterial colonies in liquid crystalline environments},
journal = {Proceedings of the National Academy of Sciences},
volume = {123},
number = {34},
pages = {e2536343123},
year = {2026},
doi = {10.1073/pnas.2536343123}}

@article{Pratten1986size,
title = {Pinocytosis and phagocytosis: the effect of size of a particulate substrate on its mode of capture by rat peritoneal macrophages cultured in vitro},
journal = {Biochimica et Biophysica Acta (BBA) - General Subjects},
volume = {881},
number = {3},
pages = {307-313},
year = {1986},
issn = {0304-4165},
doi = {https://doi.org/10.1016/0304-4165(86)90020-6},
url = {https://www.sciencedirect.com/science/article/pii/0304416586900206},
author = {Margaret K. Pratten and John B. Lloyd}
}

@article{Champion2006shape,
author = {Julie A. Champion  and Samir Mitragotri },
title = {Role of target geometry in phagocytosis},
journal = {Proceedings of the National Academy of Sciences},
volume = {103},
number = {13},
pages = {4930-4934},
year = {2006},
doi = {10.1073/pnas.0600997103}}

@article{Guttenplan2013lossofmotility,
    author = {Guttenplan, Sarah B. and Kearns, Daniel B.},
    title = {Regulation of flagellar motility during biofilm formation},
    journal = {FEMS Microbiology Reviews},
    volume = {37},
    number = {6},
    pages = {849-871},
    year = {2013},
    month = {11},
    issn = {0168-6445},
    doi = {10.1111/1574-6976.12018}
}

@article{Mohaghegh2026motility,
author = {Mohaghegh, Fatemeh and Scadden, Jacob and Baker, Matthew A. B.},
title = {To Move or Not to Move: When and How Bacteria Suppress Flagellar Motility},
journal = {Molecular Microbiology},
volume = {126},
number = {2},
pages = {173-189},
doi = {https://doi.org/10.1111/mmi.70085},
year = {2026}
}

@article{Puorger2011fimAfolding,
title = {Structure, Folding and Stability of FimA, the Main Structural Subunit of Type 1 Pili from Uropathogenic Escherichia coli Strains},
journal = {Journal of Molecular Biology},
volume = {412},
number = {3},
pages = {520-535},
year = {2011},
issn = {0022-2836},
doi = {https://doi.org/10.1016/j.jmb.2011.07.044},
url = {https://www.sciencedirect.com/science/article/pii/S0022283611008126},
author = {Chasper Puorger and Michael Vetsch and Gerhard Wider and Rudi Glockshuber}
}

\end{document}

\preprint{APS/123-QED}

\title{Supplemental Material: Fimbriation alters bacterial colony morphogenesis in polymeric environments}

\author{Udayanidhi {Ramesh Kumar}}
\thanks{These authors contributed equally to this work.}
\affiliation{William A. Brookshire Department of Chemical and Biomolecular Engineering, University of Houston, Houston, TX, USA}

\author{Tanumoy Dhar}
\thanks{These authors contributed equally to this work.}
\affiliation{Division of Chemistry and Chemical Engineering, California Institute of Technology, Pasadena, CA, USA}

\author{Sebastian {Gonzalez La Corte}}
\affiliation{Lewis-Sigler Institute for Integrative Genomics, Princeton University, Princeton, NJ, USA}
\thanks{Current address: Department of Molecular, Cellular, and Developmental Biology, Yale University, New Haven, CT, USA.}

\author{Sujit S. Datta}
\email{ssdatta@caltech.edu}
\affiliation{Division of Chemistry and Chemical Engineering, California Institute of Technology, Pasadena, CA, USA}

\author{Patrick C. Cirino}%
 \email{pccirino@uh.edu}
\affiliation{William A. Brookshire Department of Chemical and Biomolecular Engineering, University of Houston, Houston, TX, USA}

\author{Jacinta C. Conrad}%
 \email{jcconrad@uh.edu}
\affiliation{William A. Brookshire Department of Chemical and Biomolecular Engineering, University of Houston, Houston, TX, USA}

\date{\today}% It is always \today, today,

\maketitle

\section{\label{appendix:strains} Strains and plasmids used in this study}

Tables \ref{table:genetics} and \ref{table:primers} respectively contain information about the strains and plasmids used in this study, and on the primers used to construct the fluorescent plasmids.

\begin{table}[H]
\caption{\label{table:genetics}Bacterial strains and plasmids used in this study.}
\begin{ruledtabular}
\begin{tabular}{lll}
& Relevant Genetics & Source \\
\hline
 \multicolumn{3}{l}{Strains}\\
 \hline
  MG1655 & \textit{F}- $\lambda$- \textit{ilvG}- \textit{rfb}-50 \textit{rph}-1 & CGSC\# 8237 \\
  MG1655$\Delta$\textit{fliC} & MG1655, \textit{fliC} deleted & This study \\
  \hline
 \multicolumn{3}{l}{Plasmids}\\
  \hline
 p(blank) & \makecell[l]{Lab plasmid pPCC2000, derived from pPCC1322 \footnotemark[1]. \\ No ORF downstream of P\textit{tac}} & \cite{RameshKumar2024} \\
 p(\textit{fim}) & \makecell[l]{Lab plasmid pPCC1401. \\ \textit{fim} operon under the control of P\textit{tac}} & \cite{McLay2018}\\
 p(blank-\textit{egfp}) & \makecell[l]{Lab plasmid pPCC2000 with constitutive \textit{egfp} expression \\ under control of P\textit{tet}} & This study \\
 p(\textit{fim}-\textit{rfp}) & \makecell[l]{Lab plasmid pPCC1401 with constitutive \textit{mCherry} expression \\ under control of P\textit{tet}} & This study \\ 
\end{tabular}
\end{ruledtabular}
\footnotetext[1]{All plasmids contain pBR322 origin, \textit{lacI}, promoter P\textit{tac}, \textit{aac} (Apr$^\mathrm{R}$).}
\end{table}

\begin{table}[H]
\caption{\label{table:primers}Primers used in this study.}
\begin{ruledtabular}
\begin{tabular}{ll}
 \multicolumn{2}{l}{Primers}\\
 \hline
fliC-P1-FW & \makecell[l]{5$'$ CGATTAACCCTGCAGCAGAGACAGGTGTAGGCTGGAGCTGCT \\ TC 3$'$} \\
fliC-P1-RV & \makecell[l]{5$'$ CAACGACTTGCAATATAGGATAACGAATCATGTAATTCCGGGG\\ATCCGTCG 3$'$} \\
fliC-P2-FW & \makecell[l]{5$'$ GTCAGTCTCAGTTAATCAGGTTACAACGATTAACCCTGCAGCA\\GAGAC 3$'$} \\
fliC-P2-RV & \makecell[l]{5$'$ GGAAACCCAATACGTAATCAACGACTTGCAATATAGGATAACG\\AATCATG 3$'$} \\
fliC-test-FW & 5$'$ GTAGGCCTGATAAGCGCAGCG 3$'$ \\
fliC-test-RV & 5$'$ GGTGCCGCCTGCTCTCCAACCGCGGTCAAACTGGG 3$'$ \\
p1234-Ptet-XmnI-FW  & 5$'$ CAATAGGAATGGGTTCAAGCCAGTATACACTCCGCTATCG 3$'$ \\
p1234-Ptet-egfp-RV & 5$'$ TGTTTCTCCTCTTTAATGAATTCGGTCA 3$'$ \\
p1234-Ptet-egfp-FW & \makecell[l]{5$'$ ACCGAATTCATTAAAGAGGAGAAACACACATGGTGAGCAAGG\\GCGAGG 3$'$} \\
p1234-egfp-BsiWI-RV & 5$'$ AGATCACGTACGAAAAAAAAAAGCCCGCACTGTCAGG 3$'$ \\
p1234-Ptet-rfp-RV & 5$'$ TTATCCTGCAACTCCCGGGTGGTCAGTGCGTCCTGCTGAT 3$'$ \\
p2107-Ptet-rfp-FW  & 5$'$ ACCCGGGAGTTGCAGGATAATC 3$'$ \\
p2107-rfp-BsiWI-RV & 5$'$ AGATCTCGTACGAAAAAAAAAAGCCCGCACTGTCAG 3$'$ \\
\end{tabular}
\end{ruledtabular}
\end{table}

\section{Growth curves}

The growth curves for the different strains (MG1655$\Delta$\textit{fliC}+plasmids) used in this study are presented in Fig.\ \ref{fig: Growth curves}. Uninduced p(\textit{fim}-\textit{rfp}) cells grow at a slower rate than p(blank-\textit{egfp}) cells. The p(\textit{fim}-\textit{rfp}) plasmid (12907 bp) is significantly larger than the p(blank-\textit{egfp}) plasmid (5259 bp) and thus imposes a higher metabolic burden due to the greater energy consumption for plasmid replication, which hinders cell growth \cite{Tsang2017, Cheah1987}. Fimbriated (induced) p(\textit{fim}-\textit{rfp}) bacteria grow more slowly than  nonfimbriated bacteria. The growth rate decreases as the concentration of IPTG is increased from 10 $\mu$M to 50 $\mu$M because the biosynthesis and assembly of fimbriae consumes significant energy and carbon \cite{Qiao2021, Wurpel2013}.

\renewcommand{\thefigure}{S\arabic{figure}}
\begin{figure}
\includegraphics{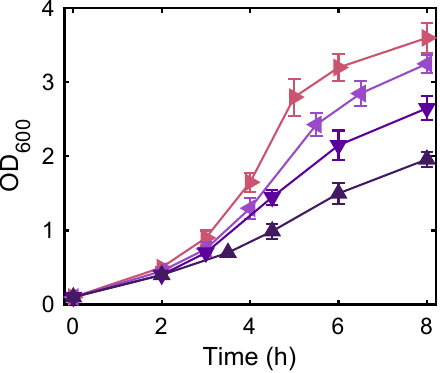}% 
\caption{\label{fig: Growth curves} Growth curves of p(blank-\textit{egfp}) ($\blacktriangleright$), uninduced p(\textit{fim}-rfp)  ($\blacktriangleleft$) and p(\textit{fim}-rfp) induced with 10 $\mu$M ($\blacktriangledown$) or 50 $\mu$M IPTG ($\blacktriangle$) in LB supplemented with 50 $\mu$g mL$^{-1}$ apramycin sulfate at 37$^\circ$C and 250 rpm. The strains were induced with 0, 10, or 50 $\mu$M IPTG at OD$_\mathrm{600}$ 0.5. Growth curves were drawn by monitoring the OD$_\mathrm{600}$ value, which represents the bacterial concentration, as a function of time. Three biological replicates were measured for each sample. Error bars represent standard deviations.}
\end{figure}

\section{Flow cytometry}

The biosynthesis and assembly of fimbriae consume significant energy and carbon \cite{Qiao2021, Wurpel2013}. As a result, inducing extremely high expression of fimbriae may affect cell viability. We confirm that cell viability is not greatly affected by induction of fimbrial expression at the inducer concentrations used in this study by staining with propidium iodide (PI) and Redox Sensor Green (RSG) dye. PI is a red fluorescent dye that does not penetrate live cells, and thus is commonly used to detect dead cells in a population \cite{Boulos1999}. Penetration of PI through the cell membrane yields high PI fluorescence, indicating that the cell is dead or membrane compromised. RSG dye penetrates the cell membrane and acts as a fluorogenic indicator of cellular redox potential and hence metabolic activity \cite{Kalyuzhnaya2008}. Low RSG fluorescence implies that the cell has reduced metabolic activity. Thus, the combination of high PI fluorescence coupled with low RSG fluorescence identifies those cells that are dead with compromised membrane activity due to fimbrial expression.

We expose p(blank) and p(\textit{fim}) cells induced with 0, 10, and 50 $\mu$M IPTG to PI and RSG dyes and examine the fluorescence signals using flow cytometry. To determine the cutoffs for the fluorescence signal levels that indicated dead and membrane-compromised cells, we conduct positive control experiments on cells treated with 70 \% v/v ethanol [Fig.\ \ref{fig: Flow cytometry plot for dead cells}]. Fewer than 1\% of p(blank) and uninduced p(\textit{fim}) cells are non-viable (i.e., dead or membrane-compromised). About 1\% and 15\% of p(\textit{fim}) cells lose their viability when grown in the presence of 10 and 50 $\mu$M IPTG, respectively. These values indicate that cell viability is not significantly affected by fimbrial expression at the IPTG concentrations used in this study.

% Figure S1: Flow cytometry plots
% \setcounter{figure}{1}
% \renewcommand{\figurename}{FIG.}
\renewcommand{\thefigure}{S\arabic{figure}}
\begin{figure*}
\includegraphics{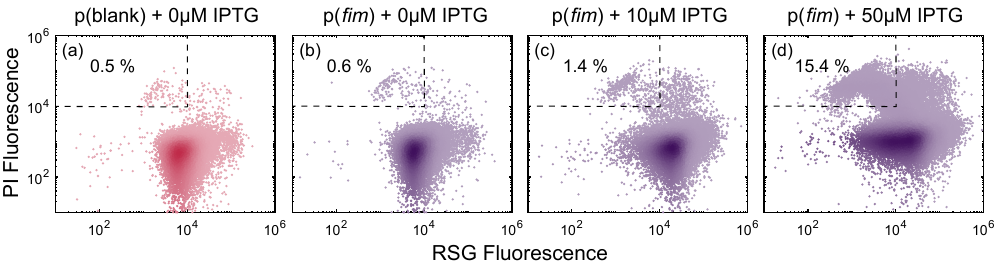}% 
\caption{\label{fig: flow cytometry plots} PI and RSG staining to determine the viability of p(blank) and p(\textit{fim}) cells induced with various IPTG concentrations used in this study. (a) p(blank) cells and (b -- d) p(\textit{fim}) cells grown in the presence of (b) 0 $\mu$M IPTG, (c) 10 $\mu$M IPTG, and (d) 50 $\mu$M IPTG for 4 h were stained with PI and RSG dye and analyzed by flow cytometry. Cell viability is not significantly affected by fimbrial expression at the IPTG concentrations used in this study. The horizontal and vertical dashed lines indicate the gating cutoffs for the PI and RSG fluorescence, respectively. The percentage values are the percentage of dead and membrane-compromised cells in the cell population.}
\end{figure*}

% Flow cytometry plot for dead cells
% \setcounter{figure}{2}
% \renewcommand{\figurename}{FIG.}
\renewcommand{\thefigure}{S\arabic{figure}}
\begin{figure}
\includegraphics{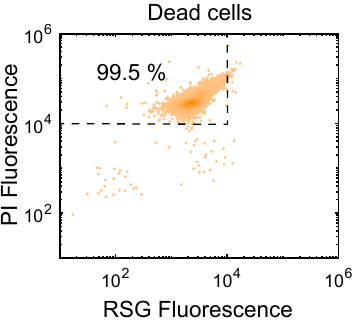}% 
\caption{\label{fig: Flow cytometry plot for dead cells} Bacteria treated with 70 \% ethanol for 1 h are stained with PI and RSG dye and analyzed by flow cytometry as a positive control to determine cell viability. The gating cutoff for both PI and RSG fluorescence was 10,000 RFU. The horizontal and vertical dashed lines indicate the gating cutoffs for the PI and RSG fluorescence, respectively. The percentage value corresponds to the percentage of dead and membrane-compromised cells in the cell population.}
\end{figure}

\section{Supplemental AFM images}

Figure \ref{fig: AFM of dividing fim cells} shows that the new ends of a dividing cell do not appear to bear fimbriae.
% AFM of dividing fim cells
% \setcounter{figure}{3}
% \renewcommand{\figurename}{FIG.}
\renewcommand{\thefigure}{S\arabic{figure}}
\begin{figure}
\includegraphics{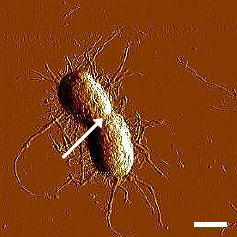}% 
\caption{\label{fig: AFM of dividing fim cells} Representative AFM image of a dividing fimbriated p(\textit{fim}) bacterium induced with 50 $\mu$M IPTG. Fimbriae do not appear to be expressed at the conjunction of two dividing cells (arrow), near the poles of the cell bodies. Scale bar is 1 $\mu$m.}
\end{figure}

\section{\label{appendix:visometry}Intrinsic viscosity measurements}

We prepare series of dilute NaPSS ($M_w = 70$ kDa) and Ficoll ($M_w = 400$ kDa) solutions in PBS and LB by diluting a concentrated stock solution of NaPSS and Ficoll in PBS and LB, respectively. The samples are homogenized using a roller mixer overnight at room temperature. The viscosities are measured as a function of NaPSS concentration using a glass capillary viscometer (Cannon Instrument Company, 0C; Fig.\ \ref{fig: Rg measurements}). To determine the intrinsic viscosity [$\eta$], the data are fitted to the two-term virial expansion of viscosity $\eta = \eta_{0} (1 + [\eta]c_\mathrm{Polymer})$, where $\eta_{0}$ and $c_\mathrm{Polymer}$ are the viscosity of the solvent and the concentration of polymer solution, respectively. From the intrinsic viscosity, we calculate the overlap concentration $c^* = 1/[\eta]$ and radius of gyration $ R_\mathrm{g,0} = (M_w /(4/3 \pi N_\mathrm{av} c^*)) ^ {1/3}$ (Table \ref{tab: Rg values}). 

\begin{table}
    \centering
\caption{Overlap concentration $c^*$ and radius of gyration $ R_g$ for NaPSS and Ficoll in PBS and LB.}
\label{tab: Rg values}
\setlength{\tabcolsep}{12pt}
    \begin{tabular}{|c|l|c|c|}\hline
 \multicolumn{2}{|c|}{}& in PBS & in LB \\\hline
 \multirow[c]{2}{*}{70 kDa NaPSS} &  $c^*$ (g L$^{-1}$)&3&14\\\cline{2-4}
         &    $R_g$ (nm) &22& 13\\\hline
 \multirow[c]{2}{*}{400 kDa Ficoll} &   $c^*$ (g L$^{-1}$)&3& 12\\\cline{2-4}
         &   $R_g$ (nm)&39& 25\\ \hline
    \end{tabular}
\setlength{\tabcolsep}{10pt}
\renewcommand{\arraystretch}{1.5}
\end{table}

% Figure S4: Rg measurements
% \setcounter{figure}{3}
% \renewcommand{\figurename}{FIG.}
\renewcommand{\thefigure}{S\arabic{figure}}
\begin{figure}
\includegraphics{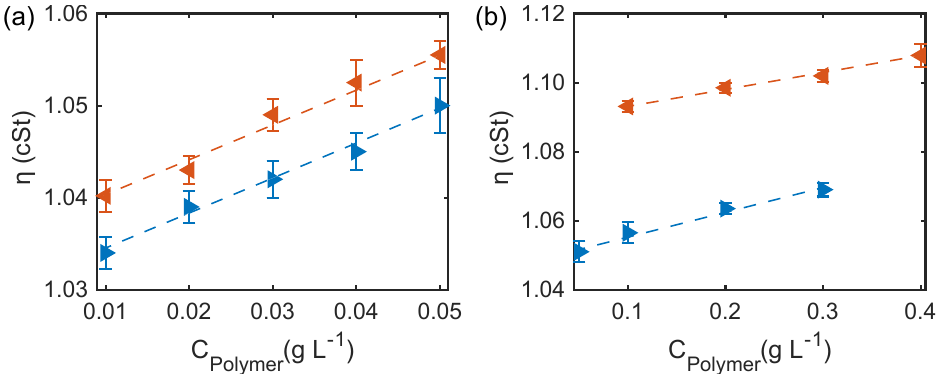}% 
\caption{\label{fig: Rg measurements} Solution viscosity $\eta$ as a function of NaPSS ($\blacktriangleright$) and Ficoll ($\blacktriangleleft$) concentration in (a) PBS and (b) LB.}
\end{figure}

\section{\label{supp:depexp} Calculation of experimental depletion potentials}

To determine the strength of the depletion-driven attraction between bacteria, we model a bacterial cell as a spherocylinder consisting of a cylinder of length \textit{l} and diameter \textit{d}, capped on each of its end by a hemisphere of the same diameter. The depletion potential between two spherocylinders \cite{savenko2006phase} is 
\begin{equation}
    U_{dep} = -\Pi_{\mathrm{osm}} \frac{\epsilon_0\sigma^2}{2}
    \left[\frac{\arccos(x) - x\sqrt{1-x^2}}
    {\sqrt{1-\chi^2(\lambda,\mu)(\mathbf n_i\cdot \mathbf n_j)^2}}\right],
    \label{eq:depletion potential}
\end{equation}
for $h_{ij}^{\min} < d + 2R_p$, where $\Pi_{\mathrm{osm}}$ is the osmotic pressure of the polymer solution, $\sigma = d + 2R_p$ is the effective depletion diameter, $R_p$ is the polymer radius, $x=h_{ij}^{\min}(\mathbf R_{ij}, \mathbf n_i,  \mathbf n_j)/\sigma$ is the minimum distance between the centerlines of spherocylinders \textit{i} and \textit{j} divided by $\sigma$, $\epsilon_0$ is a fitting parameter, $\mathbf n_i$ and $\mathbf n_j$ are corresponding unit vectors denoting the orientations of the spherocylinders \textit{i} and \textit{j}, and $\chi$ is the anisotropy parameter. This parameter satisfies  
\begin{equation}
    \chi^2(\lambda,\mu) = 
    \left(\frac{(\xi \lambda +\sigma)^{2}-\sigma^{2}}{(\xi \lambda +\sigma)^{2}+\sigma^{2}}\right)
    \left(\frac{(\xi \mu +\sigma)^{2}-\sigma^{2}}{(\xi \mu +\sigma)^{2}+\sigma^{2}}\right),
    \label{eq:anisotropy parameter}
\end{equation}
where $\xi$ is an empirical fitting parameter and $\lambda$ and $\mu$ are interaction lengths defined in Ref.~\cite{savenko2006phase}. The equation \ref{eq:depletion potential} vanishes when $h_{ij}^{\min} \ge d + 2R_p$. 

When two spherocylinders have end-to-end depletion attractions, the cylindrical portions do not overlap in the excluded volume and only the two hemispherical ends interact. Thus, the end-to-end overlapping excluded volume between two spherocylinders of diameter \textit{d} is equivalent to the end-to-end overlapping excluded volume between two spheres with the same diameter. The end-to-end depletion potential between two spherocylinders of diameter \textit{d}, separated by a distance \textit{r} \cite{verma1998entropic} is thus
\begin{equation}
    U_{ee} = -\Pi_{\mathrm{osm}} \frac{\pi}{4}
    \left[d\lambda(r-d\lambda)^2 + \frac{1}{3} (r-d\lambda)^3\right]
    \label{eq: end-to-end depletion potential}
\end{equation}
for $d \leq r \leq d + 2Rp$, where $\lambda = 1 + 2Rp/d$. Equation \ref{eq: end-to-end depletion potential} is valid when the polymer concentration is less than the overlap concentration $c ^ \ast $. 

When the polymer concentration is greater than $c ^ \ast $, the end-to-end depletion potential is predicted (with the Derjaguin approximation, i.e., $\sigma \gg \pi\xi$) to be \cite{verma1998entropic}
\begin{equation}
    U_{ee} \approx -\Pi_{\mathrm{osm}} \frac{\pi d}{4}
    \left[r-(d+\pi\xi)\right]^2
\end{equation}
for $d \leq r \leq d + \pi\xi$, where $\xi = (2/\pi)R_p$ is the correlation length.

\section{\label{appendix:confocal micrographs at non-growth conditions} Confocal micrographs of bacteria in nutrient-poor polymer solutions}

% Figure S5: confocal micrographs at non-growth conditions
% \setcounter{figure}{4}
% \renewcommand{\figurename}{FIG.}
\renewcommand{\thefigure}{S\arabic{figure}}
\begin{figure}[H]
\includegraphics{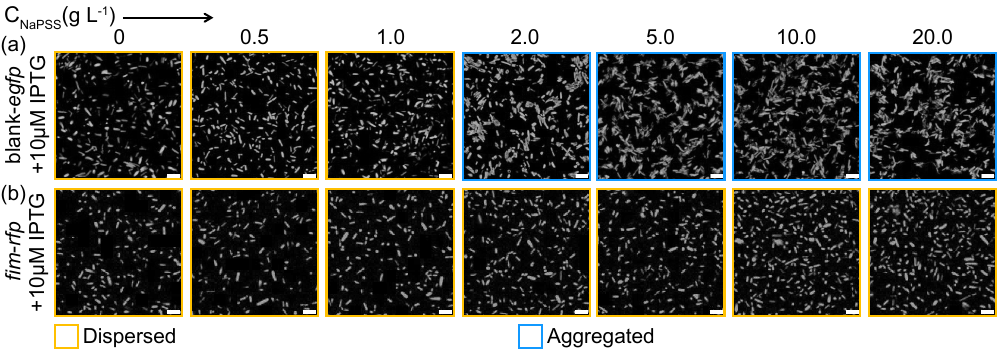}% 
\caption{\label{fig: confocal micrographs at non-growth conditions_appendix} (Color online) Confocal micrographs (colored in grayscale) of (a) p(blank-\textit{egfp}) and (b) p(\textit{fim}-\textit{rfp}) cells induced with 10 $\mu$M IPTG in PBS containing NaPSS that are acquired 1 h after addition of polymer. p(blank-\textit{egfp}) cells induced with 10 $\mu$M IPTG aggregate side-by-side when [NaPSS] $\geq  5$ g L$^{-1}$ ($U_\mathrm{ss} \geq 143$ k$_\mathrm{B}$T). By contrast, p(\textit{fim}-\textit{rfp}) cells induced with 10 $\mu$M IPTG remain dispersed for NaPSS concentrations of up to 20 g L$^{-1}$. Scale bar is 5 $\mu$m.}
\end{figure}

\section{\label{appendix:ficoll} Bacteria in Ficoll solutions}

Ficoll is a neutral and highly branched polymer. Figure \ref{fig: Ficoll state diagram} shows that the sedimentation state diagrams for non-fimbriated (p(blank)) and fimbriated (p(\textit{fim})) induced with 10 $\mu$M IPTG) bacteria suspended in solutions of Ficoll in PBS are qualitatively similar to those obtained for bacteria in NaPSS solutions. 

% Figure S6: State diagrams for Ficoll
% \setcounter{figure}{5}
% \renewcommand{\figurename}{FIG.}
\renewcommand{\thefigure}{S\arabic{figure}}
\begin{figure}[H]
\centering
\includegraphics{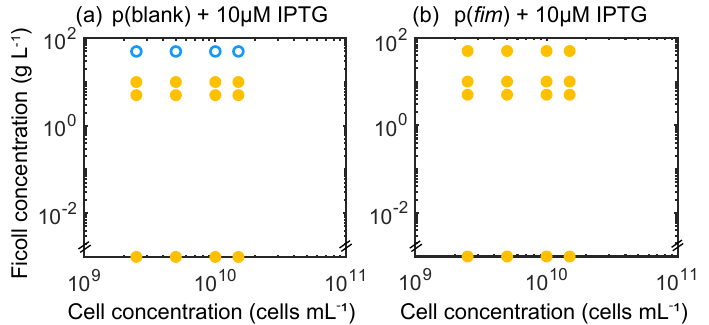}% 
\caption{\label{fig: Ficoll state diagram} Fimbriation inhibits Ficoll-induced depletion aggregation of \textit{E.\ coli} under non-growth conditions. State diagram for (a) p(blank) and (b) p(\textit{fim}) cells induced with 10 $\mu$M IPTG in PBS with Ficoll. Open circles indicate aggregated suspensions and filled circles indicate dispersed cell suspensions, determined visually 4 h after addition of polymer. p(blank) but not p(\textit{fim}) cells aggregate at Ficoll concentrations of 50 g L$^{-1}$.}
\end{figure}

\newpage

Figure \ref{fig: Cables and strings in Ficoll} shows that non-fimbriated (p(blank-\textit{egfp})) and fimbriated (p({\textit{fim}-\textit{rfp})) bacteria grown in Ficoll solutions (in LB) respectively form cables and strings that are qualitatively similar to those formed by the bacteria in NaPSS solutions (in LB).

% Figure S7: Cables and strings in Ficoll solution
% \setcounter{figure}{6}
% \renewcommand{\figurename}{FIG.}
\vspace{-0.1in}
\renewcommand{\thefigure}{S\arabic{figure}}
\begin{figure}[H]
\centering
\includegraphics{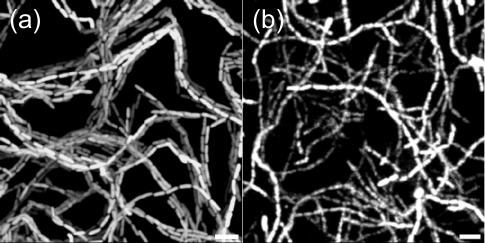}% 
\caption{\label{fig: Cables and strings in Ficoll} Non-fimbriated bacteria form cables and fimbriated bacteria form strings in nutrient-rich Ficoll solutions. 2-D projections of 3-D confocal micrographs (colored in grayscale) of (a) p(blank-\textit{egfp}) and p(\textit{fim}-\textit{rfp}) cells after 5 h of growth in LB medium containing 50 g L$^{-1}$ Ficoll and 50 $\mu$M IPTG. Scale bar is 5 $\mu$m.}
\end{figure}

% Figure S8: Transition of strings to cables at 10uM IPTG
% \setcounter{figure}{7}
% \renewcommand{\figurename}{FIG.}
\vspace{-0.1in}
\renewcommand{\thefigure}{S\arabic{figure}}
\begin{figure}[H]
\centering
\includegraphics{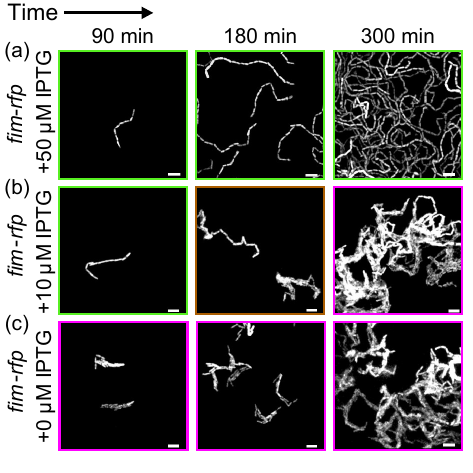}% 
\caption{\label{fig: strings to cables} (Color online) Transition from strings to cables in LB medium containing 20 g L$^{-1}$ NaPSS and 10 $\mu$M IPTG. (a -- c) 2-D projections of 3-D confocal micrographs (grayscale) of p(\textit{fim}-\textit{rfp}) cells induced with (a) 50 $\mu$M IPTG, (b) 10 $\mu$M IPTG, and (c) 0 $\mu$M IPTG in LB medium containing 20 g L$^{-1}$ NaPSS at different time points indicated in the figure. Micrographs outlined in magenta, green, or brown, respectively, indicate cables, strings, or transition state. Scale bar is 5 $\mu$m.}
\end{figure}

% Figure S10: Transition of strings to cables at 10uM IPTG
% \setcounter{figure}{8}
% \renewcommand{\figurename}{FIG.}
\renewcommand{\thefigure}{S\arabic{figure}}
\begin{figure}[H]
\centering
\includegraphics{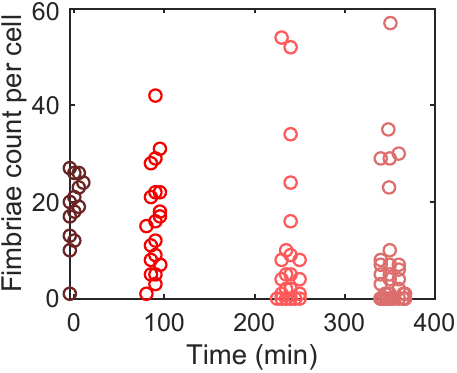}% 
\caption{\label{fig: fimbriation vs time} Number of p(\textit{fim}-\textit{rfp}) cells expressing fimbriae, as determined using AFM measurements, as a function of time for bacteria grown in the presence of 10 $\mu$M IPTG.  Fimbrial expression is more heterogeneous on long time scales and fewer bacteria bear fimbriae under this induction condition.}
\end{figure}

\section{\label{appendix:simulations}Agent-based simulations}

We perform two-dimensional agent-based Brownian dynamics simulations of proliferating bacteria, extending the model developed previously to describe cable formation of non-fimbriated cells~\cite{GonzalezLaCorte2025} with the addition of a fimbria-induced steric repulsion. The model is summarized schematically in Fig.~\ref{fig: model schematic}. 

\begin{figure}[H]
\centering
\includegraphics[width=0.35\textwidth]{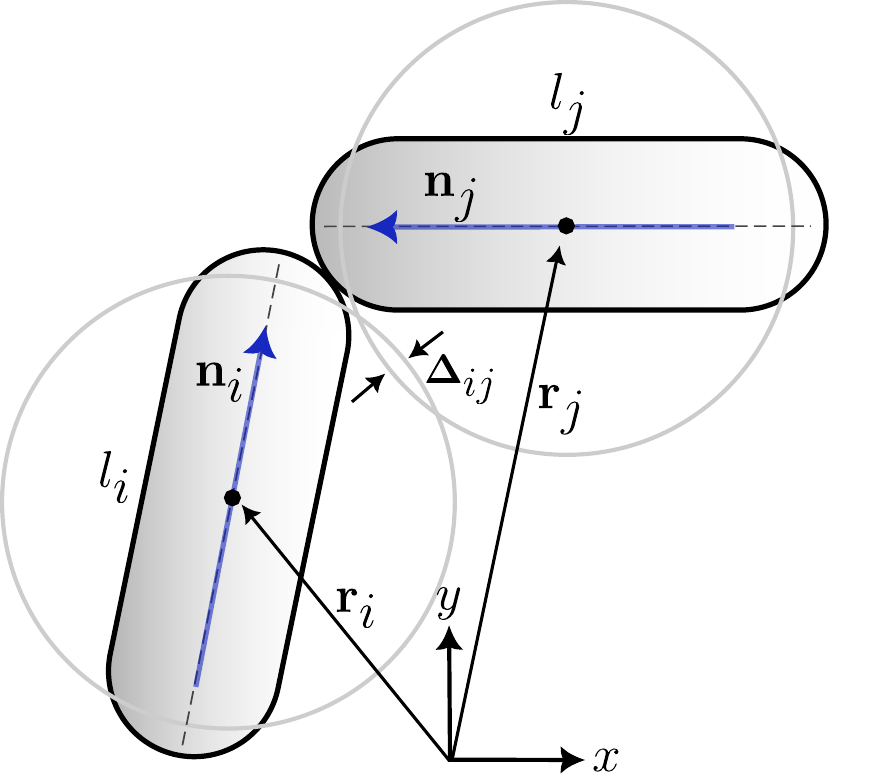}
\caption{\label{fig: model schematic} Schematic of the agent-based model. Cells are modeled as spherocylinders with cylindrical body length $l$ and hemispherical end caps of diameter $d$. The unit vector $\mathbf{n}$ denotes the cell orientation. Fimbriae are represented as a soft exclusion layer that produces an additional repulsive interaction when fimbriated regions overlap.}
\end{figure}

\subsection{Mechanics of growth and division}

We model each bacterium as a spherocylinder composed of a cylindrical body of length $l_i$ and two hemispherical end caps of diameter $d$. The total cell volume is
\begin{equation}
    V_i(t) = \frac{\pi d^3}{6} + \frac{\pi d^2 l_i(t)}{4},
\end{equation}
where the first term is the volume of the two hemispherical caps and the second term is the volume of the cylindrical body. Cell growth occurs by elongation at fixed diameter. Assuming exponential volume growth with rate $\Gamma_i$, $dV_i/dt = \Gamma_i V_i$, the cylindrical body length evolves as
\begin{equation}
    l_i(t) = e^{\Gamma_i t}\left(l_0+\frac{2d}{3}\right) - \frac{2d}{3},
    \label{eq:growth_dimensional}
\end{equation}
where $l_0$ is the initial cylindrical body length and $\Gamma_i$ is the growth rate of cell $i$. Cells divide when the body 
length reaches $l_i(t_d) = 2l_0+d$, giving the division time
\begin{equation}
        t_d = \frac{1}{\Gamma}\ln\!\left(\frac{5d + 6l_0}{2d + 3l_0}\right)
        = \frac{\ln(17/8)}{\Gamma} \approx 0.754\,\Gamma^{-1},
\end{equation}
where $\Gamma$ is the mean rate. Upon division, the mother cell is replaced by two daughter cells of equal length. This division rule does not exactly conserve cell volume: replacing a mother of body length $2l_0+d$ by two daughters of body length $l_0$ removes a volume $\pi d^3/12$, because the discarded cylindrical segment of length $d$ exceeds the newly formed pair of hemispherical caps. The loss is a fixed increment per division, independent of $l_0$, and does not accumulate with cell size; at the simulated values $d = 1$ and $l_0 = 2d$, it amounts to $1/17 \approx 5.9\%$ of the dividing cell's volume. We retain this rule because it fixes the daughter length to $l_0$, and the per-division offset does not affect the collective mechanics that are the subject of this work. We report lengths in units of $d$ and times in units of $\Gamma^{-1}$, so that $l_0 = 2$ and $t_d \approx 0.754$. At division, each daughter draws its growth rate $\Gamma_i$ independently of its mother's from a Gaussian distribution with mean $\Gamma$ and standard deviation $0.2\,\Gamma$. This spread desynchronizes division, commonly used in the models of growing colonies~\cite{you2018geometry}. Since the birth and division lengths are both fixed, $\Gamma_i$ is the only stochastic element of the cell cycle: the interdivision time $\tau_{\mathrm{div},i} = \ln(17/8)/\Gamma_i$ inherits all of its variability from $\Gamma_i$.

\subsection{Cell--cell elastic repulsion}

Steric repulsion between overlapping cells is modeled using a Hertzian contact energy. Let $r_{ij}^{\min}$ denote the shortest distance between the two spherocylinder axes, so that the overlap is $\delta_{ij} = d - r_{ij}^{\min}$. The elastic interaction energy is
\begin{equation}
    \phi_{ij}^{\mathrm{el}} =
    \begin{cases}
    \dfrac{1}{\sqrt{2}}\,E_0 d^{1/2}\delta_{ij}^{5/2}, & \delta_{ij}>0,\\[6pt]
    0, & \delta_{ij}\le 0,
    \end{cases}
    \label{eq:elastic_energy}
\end{equation}
where $E_0$ is the effective cell stiffness. The corresponding force and torque are  obtained from gradients of $\phi_{ij}^{\mathrm{el}}$ with respect to position and orientation.

\subsection{Polymer-induced depletion attraction}
\label{sub:depletion}

Polymer depletants of radius of gyration $R_p$ generate an effective attraction between nearby cells when their polymer-exclusion shells overlap. Defining the effective depletion diameter $\sigma = d + 2R_p$ and the scaled separation $r = r_{ij}^{\min}/\sigma$, depletion acts for $r<1$, with potential
\begin{equation}
    \phi_{ij}^{\mathrm{dep}} =
    \begin{cases}
    -\Pi^{\mathrm{osm}} V_{ij}^{\mathrm{ev}}, & r<1,\\[6pt]
    0, & r\ge 1,
    \end{cases}
\end{equation}
where $\Pi^{\mathrm{osm}}$ is the osmotic pressure of the polymer solution and $V_{ij}^{\mathrm{ev}}$ is the overlap of the excluded-volume regions. Following Refs.~\cite{savenko2006phase,GonzalezLaCorte2025},
\begin{equation}
    V_{ij}^{\mathrm{ev}} = \frac{\epsilon_0\sigma^2}{2}
    \left[\frac{\cos^{-1}(r) - r\sqrt{1-r^2}}
    {\sqrt{1-\chi^2(\lambda,\mu)(\mathbf n_i\cdot \mathbf n_j)^2}}\right],
    \label{eq:excluded_volume}
\end{equation}
with $\epsilon_0$ a geometric fitting parameter and the orientation unit vector $\mathbf{n}_i = (\cos\theta_i,\sin\theta_i)$ as defined in Fig.~\ref{fig: model schematic}. The anisotropy factor is
\begin{equation}
    \chi^2(\lambda,\mu) = 
    \left(\frac{(\xi \lambda +\sigma)^{2}-\sigma^{2}}{(\xi \lambda +\sigma)^{2}+\sigma^{2}}\right)
    \left(\frac{(\xi \mu +\sigma)^{2}-\sigma^{2}}{(\xi \mu +\sigma)^{2}+\sigma^{2}}\right),
\end{equation}
where $\xi$ is a second empirical fitting parameter and $\lambda$ and $\mu$ are interaction lengths defined in Ref.~\cite{savenko2006phase}. The values of $\epsilon_0$ and $\xi$ are taken from Ref.~\cite{GonzalezLaCorte2025}. In the  two-dimensional implementation used here, $V_{ij}^{\mathrm{ev}}$ is interpreted as the corresponding excluded-area overlap, with $\epsilon_0$ absorbing the effective out-of-plane length scale.

The lengths $\lambda$ and $\mu$ measure the axial extent over which the surface of one cell lies within $\sigma$ of the other, and they are what distinguish the two contact configurations. For two cells of body length $l$ in side-by-side contact the bodies share the same axial span, so $\lambda = \mu = l$ and $\chi^2 \to 1$, maximizing the attraction. For two collinear cells in pole-to-pole contact the bodies occupy disjoint intervals along the shared axis, so $\lambda = \mu = 0$ and $\chi^2 = 0$: the poles receive the bare overlap kernel with no orientational enhancement. The depletion attraction is therefore anisotropic, favoring side-by-side over end-to-end contact.

\subsection{Fimbria-induced soft repulsion}

Fimbriae distributed over the cell surface generate an additional soft repulsive interaction when the fimbriated exclusion regions of neighboring cells overlap. We model the fimbrial exclusion of each cell $i$ as a sphere of radius $l_i/2 + fd/2$ centered on the cell, with $f = 0.65$. Because the pole tip sits at $l_i/2 + d/2$, each bare pole protrudes a distance $(1-f)d/2 = 0.175\,d$ beyond the exclusion surface, while the lateral surface remains coated. The exclusion is therefore anisotropic in effect---bald poles, coated flanks---even though its geometry is isotropic.

Two cells first exclude when their spheres touch, at center-to-center distance
\begin{equation}
    R_{ij}^{\mathrm{fm}} = \frac{l_i+l_j}{2}+fd .
\end{equation}
Side-by-side, two cells of length $l$ contact at center separation $d$, where their spheres overlap by $l+(f-1)d \approx l$---an exclusion of order the full cell length. End-to-end and collinear, the hard poles collide at center separation $l+d$, whereas the exclusion spheres would meet only at the smaller separation $l+fd$; the poles therefore touch with the shells still $(1-f)d = 0.35\,d$ apart, so fimbrial exclusion never engages pole-to-pole. Cells thus pay an exclusion of order $l$ to sit side-by-side but none to sit pole-to-pole.

The scalar overlap of the fimbriated regions is
\begin{equation}
    \Delta_{ij}^{\mathrm{fm}} = R_{ij}^{\mathrm{fm}} - |\mathbf r_i-\mathbf r_j|,
    \label{eq:fim_overlap_linear}
\end{equation}
and the linear fimbrial force on cell $i$ due to cell $j$ is
\begin{equation}
    \mathbf F_{ij}^{\mathrm{fm}} =
    \begin{cases}
    \mathcal K\,\Delta_{ij}^{\mathrm{fm}}\,\hat{\mathbf r}_{ij}, 
    & \Delta_{ij}^{\mathrm{fm}}>0,\\[6pt]
    \mathbf 0, & \Delta_{ij}^{\mathrm{fm}}\le 0,
    \end{cases}
    \label{eq:fim_force_linear}
\end{equation}
where $\mathcal K$ is the effective linear stiffness of the fimbriated layer, with 
units of force per length, and $\hat{\mathbf r}_{ij} = (\mathbf r_i-\mathbf r_j)/
|\mathbf r_i-\mathbf r_j|$. Because the exclusion region is centered on the cell, 
the resulting force is central and exerts no torque.

\paragraph{Choice of $f$.} The parameter $f$ sets how far the exclusion sphere extends toward the cell poles, and therefore the effective bending rigidity of a string: at fixed intercellular angle, increasing $f$ increases the overlap of the pole-region exclusion spheres and generates a stronger restoring force toward a straight configuration. We fix $f = 0.65$ throughout, chosen so that the apparent persistence of simulated strings qualitatively matches that observed experimentally. The comparison is not a quantitative fit: the persistence length was not measured directly, and $f$ was selected on the basis of visual agreement. Fimbriation in the simulations is therefore varied through the bristle-layer stiffness $\mathcal K$ at fixed exclusion geometry.

\paragraph{Order-of-magnitude estimate of $\mathcal K$.} We treat the fimbriated surface as an elastic ``bristle layer'': an array of slender filaments that collectively resists deformation. Each fimbria acts as an elastic beam of effective modulus $E_f$, length $L_f$, and radius $a$, so that a single filament of cross-sectional area $A = \pi a^2$ carries a stretch stiffness $k = E_f A/L_f = \pi E_f a^2/L_f$. Summing over the $\rho$ fimbriae anchored on the cell gives
\begin{equation}
    \mathcal K = \rho\,\frac{\pi E_f a^2}{L_f}.
    \label{eq:fim_linear_stiffness}
\end{equation}
Here $E_f$ is an effective modulus capturing the filament's force--extension response rather than the intrinsic Young's modulus of the pilus rod, since fimbriae elongate largely by uncoiling their helical quaternary structure. Evaluating Eq.~\eqref{eq:fim_linear_stiffness} at the highest induction level, 50 $\mu$M IPTG, with the parameters in Table~\ref{tab: sim params} gives $\mathcal K \approx 800$ kPa$\cdot\mu$m (for $L_f = 0.2\;\mu$m); taking instead the longest fimbriae observed, $L_f = 1.6\;\mu$m, gives $\mathcal K \approx 100$ kPa$\cdot\mu$m. We emphasize that this estimate treats each fimbria as axially loaded and therefore provides only an upper bound on the layer stiffness; fimbriae in fact deform far more readily by uncoiling and by pivoting at their membrane anchor. The simulations operate well below this bound, and $\mathcal K$ is swept parametrically rather than fixed at the estimated value.

\paragraph{Fimbrial reach $l_f$.} The exclusion sphere of cell $i$ is centered on its centroid and has radius $l_i/2 + fd/2$; across the long axis, it therefore reaches beyond the cell surface by $\left[l_i + (f-1)d\right]/2$. We define the fimbrial length as this lateral reach for a newborn cell,
\begin{equation}
\ell_f \equiv \frac{l_0}{2} + \frac{fd}{2} - \frac{d}{2}
       = \frac{l_0 + (f-1)\,d}{2} = 0.825\,d ,
\label{eq:ellf}
\end{equation}
with $l_0 = 2d$ and $f = 0.65$. For $d = 1$ $\mu\mathrm{m}$, $\ell_f = 0.825$ $\mu\mathrm{m}$, within the $0.2$--$1.6$ $\mu\mathrm{m}$ range of lengths of type~1 fimbria (Fig.~\ref{M-fig:AFM images of strains used in this study}).

\subsection{Governing equations}

In the overdamped limit, the center-of-mass position $x_i^\alpha$ evolves as
\begin{equation}
    \frac{d x_i^\alpha}{dt} = \frac{1}{\eta d}\mathcal M_i^{\alpha\beta}F_i^\beta
    + \sqrt{\frac{2k_{\mathrm B}T}{\eta d}}\,\mathcal L_i^{\alpha\beta}\xi_i^\beta(t),
\end{equation}
and the orientation angle $\theta_i$ as
\begin{equation}
    \frac{d\theta_i}{dt} = \frac{\mathcal M_i^\theta}{\eta d^3}T_i
    + \sqrt{\frac{2k_{\mathrm B}T\mathcal M_i^\theta}{\eta d^3}}\,\xi_i^\theta(t),
\end{equation}
where repeated Greek indices are summed over Cartesian components, $\eta$ is the dynamic viscosity, $\mathcal M_i^{\alpha\beta}$ and $\mathcal M_i^\theta$ are the translational and rotational mobilities, and $\mathcal L_i^{\alpha\gamma} \mathcal L_i^{\beta\gamma} = \mathcal M_i^{\alpha\beta}$. The translational mobility tensor is
\begin{equation}
    \mathcal M_i^{\alpha\beta} = M_{\parallel,i}n_i^\alpha n_i^\beta
    + M_{\perp,i}\left(\delta^{\alpha\beta} - n_i^\alpha n_i^\beta\right),
\end{equation}
with the parallel, perpendicular, and rotational mobilities given by~\cite{kim2013microhydrodynamics,lobo2021polymer,shimizu1962effect}
% \CJCC{261022: these citations are broken}
\begin{equation}
    M_{\parallel,i} = \frac{\left[\left(2\mathcal A_i^2-\mathcal B^2\right)\mathcal{S}_i-2\mathcal A_i\right]\mathcal{B}}{8\pi\left(\mathcal A_i^2-\mathcal B^2\right)},
    \quad
    M_{\perp,i} = \frac{\left[\left(2\mathcal A_i^2-3\mathcal B^2\right)\mathcal{S}_i+2\mathcal A_i\right]\mathcal{B}}{16\pi\left(\mathcal A_i^2-\mathcal B^2\right)},
    \nonumber
\end{equation}
\begin{equation}
    \mathcal{M}_i^\theta = \frac{\left[3\left(2\mathcal A_i^2-\mathcal B^2\right)\mathcal{S}_i-2\mathcal A_i\right]\mathcal{B}}{16\pi\left(\mathcal A_i^4-\mathcal B^4\right)},
\end{equation}
where $\mathcal A_i = (l_i+d)/2$, $\mathcal B = d/2$, and the shape factor is
\begin{equation}
    \mathcal{S}_i = \frac{2}{\sqrt{\mathcal A_i^2-\mathcal B^2}}
    \ln\!\left(\frac{\mathcal A_i+\sqrt{\mathcal A_i^2-\mathcal B^2}}{\mathcal B}\right).
\end{equation}
The translational and rotational noises are independent Gaussian white noises with zero mean and correlations $\langle\xi_i^\alpha(t)\xi_j^\beta(t')\rangle = 
\delta_{ij}\delta^{\alpha\beta}\delta(t-t')$ and 
$\langle\xi_i^\theta(t)\xi_j^\theta(t')\rangle = \delta_{ij}\delta(t-t')$. The net 
force on cell $i$ is $F_i^\alpha = \sum_{j\ne i}(F_{ij}^{\mathrm{el},\alpha} + F_{ij}^{\mathrm{dep},\alpha} + F_{ij}^{\mathrm{fm},\alpha})$, and the net torque is $T_i = \sum_{j\ne i}(T_{ij}^{\mathrm{el}} + T_{ij}^{\mathrm{dep}})$.

\subsection{Non-dimensionalization}

We nondimensionalize lengths by the cell diameter $d$ and time by the inverse growth rate $\Gamma^{-1}$, so that $\tilde x_i^\alpha = x_i^\alpha/d$, $\tilde l_i = l_i/d$, and $\tilde t = \Gamma t$. The dimensionless growth law follows from Eq.~\eqref{eq:growth_dimensional} as $d\tilde l_i/d\tilde t = \tilde l_i + 2/3$. The dimensionless equations of motion are
\begin{equation}
\begin{split}
    \frac{d\tilde x_i^\alpha}{d\tilde t} = \;
    &-\mathcal C_{\mathrm{diff}}\mathcal C_{\mathrm{el}}\mathcal M_i^{\alpha\beta}
    \sum_{j\ne i}\frac{\partial \tilde\delta_{ij}^{5/2}}{\partial \tilde x_i^\beta}
    + \mathcal C_{\mathrm{diff}}\mathcal C_{\mathrm{dep}}\mathcal M_i^{\alpha\beta}
    \sum_{j\ne i}\frac{\partial \tilde V_{ij}^{\mathrm{ev}}}{\partial \tilde x_i^\beta}\\
    &+ \mathcal C_{\mathrm{diff}}\mathcal C_{\mathrm{fim}}\mathcal M_i^{\alpha\beta}
    \sum_{j\ne i}\tilde F_{ij}^{\mathrm{fm},\beta}
    + \sqrt{2\mathcal C_{\mathrm{diff}}}\,\mathcal L_i^{\alpha\beta}\tilde\xi_i^\beta(\tilde t),
\end{split}
\end{equation}
\begin{equation}
\begin{split}
    \frac{d\theta_i}{d\tilde t} = \;
    &-\mathcal C_{\mathrm{diff}}\mathcal C_{\mathrm{el}}\mathcal M_i^\theta
    \sum_{j\ne i}\frac{\partial \tilde\delta_{ij}^{5/2}}{\partial \theta_i}
    + \mathcal C_{\mathrm{diff}}\mathcal C_{\mathrm{dep}}\mathcal M_i^\theta
    \sum_{j\ne i}\frac{\partial \tilde V_{ij}^{\mathrm{ev}}}{\partial \theta_i}
    + \sqrt{2\mathcal C_{\mathrm{diff}}\mathcal M_i^\theta}\,\tilde\xi_i^\theta(\tilde t),
\end{split}
\end{equation}
where the dimensionless groups are
\begin{equation}
    \mathcal C_{\mathrm{el}} = \frac{E_0 d^3}{\sqrt{2}\,k_{\mathrm B}T},
    \qquad
    \mathcal C_{\mathrm{diff}} = \frac{k_{\mathrm B}T}{d^3\eta\Gamma},
    \qquad
    \mathcal C_{\mathrm{dep}} = \frac{\Pi^{\mathrm{osm}}d^3}{k_{\mathrm B}T},
    \qquad
    \mathcal C_{\mathrm{fim}} = \frac{\mathcal K d^2}{k_{\mathrm B}T}.
\end{equation}
Here $\mathcal C_{\mathrm{el}}$ compares the elastic contact energy to the thermal energy; $\mathcal C_{\mathrm{diff}}$ compares the growth time scale to the Brownian diffusion time scale, and is related to the ratio of rotational diffusion time to doubling time, $t_{\mathrm{rot}}/t_d = 1.33 (\mathcal M_i^\theta C_{\mathrm{diff}})^{-1}$, used in the main text; $\mathcal C_{\mathrm{dep}}$ measures the strength of the depletion attraction relative to thermal energy, and sets $U_{\mathrm{ee}}/k_{\mathrm B}T$; and $\mathcal C_{\mathrm{fim}}$ measures the stiffness of the fimbrial layer relative to thermal energy, and sets $\mathcal K\ell_f^2/k_{\mathrm B}T$. Because the simulations are performed entirely in reduced units, the viscosity $\eta$ and growth rate $\Gamma$ enter only through these groups.

\subsection{Simulation protocol and parameters}

Each simulation is initialized with a single cell whose position and orientation are randomly sampled within the computational domain. We consider a square computational domain of size 100$d$ $\times$100$d$ ($d=1$ $\mu\mathrm{m}$), with periodic boundary conditions imposed along both spatial ($x,y$) directions. We integrate the overdamped Langevin equations for the position and orientation of each cell with an explicit Euler--Maruyama scheme with adaptive step size~\cite{GonzalezLaCorte2025}.  Each step starts from $\Delta t_0 = 10^{-8}\,t_g \approx 7.5\times10^{-9}\,\Gamma^{-1}$, where $t_g = \Gamma^{-1}\ln(17/8) \approx 0.754\,\Gamma^{-1}$ is the doubling time at the mean growth rate ($\Gamma$). If the trial step would translate any cell's centroid by more than $0.01\,d$, we halve the step for all cells and redraw every increment, repeating until no centroid exceeds this bound. All cells therefore advance with a common step $\Delta t = \Delta t_0/2^{m} \le \Delta t_0$, where the integer $m \ge 0$ counts the halvings needed in that step, and no centroid moves farther than $0.01\,d$ per step, one-fifteenth of the depletion range $2R_p = 0.15\,d$. Throughout, we use $\mathcal{C}_\mathrm{el} = E_0 d^{3} (\sqrt{2}\,k_\mathrm{B}T)^{-1} = 5.5\times10^{4}$ and $\mathcal{C}_\mathrm{diff} = k_\mathrm{B}T (\eta\Gamma d^{3})^{-1} = 500$, and we vary $U_{\mathrm{ee}}/k_{\mathrm B}T$ and $\mathcal{K}\ell_f^{2}/k_{\mathrm B}T$ over several decades. Three independent simulations were performed at each point in the state diagram of Fig.~\ref{M-fig:simulations}(c). The dimensional parameters used are 
listed in Table~\ref{tab: sim params}.

\begin{table}[H]
    \centering
    \caption{\label{tab: sim params} Dimensional parameters used in the agent-based simulations. Upper block: parameters governing cell mechanics and polymer depletion. Lower block: fimbria-specific parameters used to estimate 
    $\mathcal K$ in Eq.~\eqref{eq:fim_linear_stiffness}.}
    % \CJCC{261022: several broken references in the lower block}
    \renewcommand{\arraystretch}{1.2}
    \begin{tabular}{l l l}
        \hline\hline
        \textbf{Parameter} & \textbf{Definition} & \textbf{Value} \\ 
        \hline 
        $d$                & Cell diameter              & $1\;\mu\mathrm{m}$~\cite{GonzalezLaCorte2025} \\  
        $l_0$              & Initial body length        & $2\;\mu\mathrm{m}$~\cite{GonzalezLaCorte2025} \\  
        $k_{\mathrm{B}}T$  & Thermal energy             & $4.14\times10^{-21}\;\mathrm{J}$~\cite{GonzalezLaCorte2025} \\  
        $E_0$              & Cell stiffness             & $325\;\mathrm{Pa}$~\cite{GonzalezLaCorte2025} \\  
        $\Pi^{\mathrm{osm}}$ & Osmotic pressure         & $0\text{--}6.67\;\mathrm{Pa}$~\cite{GonzalezLaCorte2025} \\          
        $R_{p}$            & Polymer radius of gyration & $75\;\mathrm{nm}$~\cite{GonzalezLaCorte2025} \\
        \hline 
        $E_{f}$            & Fimbria effective modulus  & $0.1\;\mathrm{GPa}$~\cite{forero2006uncoiling} \\
        $l_f$              & Lateral reach of the exclusion region at birth        & $0.825 \;\mu\mathrm{m}$~\cite{forero2006uncoiling,berne2015adhesins,Hospenthal2017} \\
        $a$                & Fimbria radius             & $3.5\;\mathrm{nm}$~\cite{Hospenthal2017} \\    
                $\rho$             & Fimbriae per cell          & $42$ [Measured, Fig.\ \ref{M-fig:AFM images of strains used in this study}] \\
        \hline\hline
    \end{tabular}
\end{table}

We emphasize that the entries in Table~\ref{tab: sim params} are the parameter  values used to run and nondimensionalize the simulations, not measured properties of the polymer used in our experiments. In particular, $R_p$ and $\Pi^{\mathrm{osm}}$ are inherited from Ref.~\cite{GonzalezLaCorte2025} and differ from the values characterizing the NaPSS solutions used here (Table~\ref{tab: Rg values}). The discrepancy does not affect our conclusions, because the dynamics depend only on the dimensionless groups defined above, and contact with experiment is made through the single lumped quantity $U_{\mathrm{ee}}/k_{\mathrm B}T$: the depletion well depth is set by the geometric prefactor $\epsilon_0$, which absorbs the overall energy scale, so a given $U_{\mathrm{ee}}/k_{\mathrm B}T$ can be reached independently of the particular $R_p$ and $\Pi^{\mathrm{osm}}$ that would produce it.

Matching $U_{\mathrm{ee}}/k_{\mathrm B}T$ constrains the end-to-end well depth but not the interaction range, which governs how the attraction is partitioned among relative orientations. In the simulations, the ratio of side-by-side to end-to-end well depths grows with cell length, from $U_{\mathrm{dep}}/U_{\mathrm{ee}} \approx 4.2$ at the newborn length ($l_i = 2d$) to $\approx 9.5$ at division ($l_i = 5d$), a range that spans the value estimated from our experiments.

We note also that the cluster state is heterogeneous: the snapshot shown in Fig.~\ref{M-fig:simulations}(d)  is representative of one type of cluster only. Because the structure of the cluster state is not the focus of the present work, we do not characterize it further; the stacked configurations reported previously in Ref.~\cite{GonzalezLaCorte2025} are also classified as clusters here.

\subsection{Sensitivity to $t_{\mathrm{rot}}/t_d$ and to the form of the fimbrial force}

Figure~\ref{fig:trot} shows string morphologies obtained at four values of $t_{\mathrm{rot}}/t_d$ spanning a decade. Varying this parameter shifts the phase boundaries but produces no qualitative change in the observed morphologies, consistent with our prior work~\cite{GonzalezLaCorte2025}; we therefore hold it fixed throughout.

\begin{figure}[H]
\centering
\includegraphics[width=\textwidth]{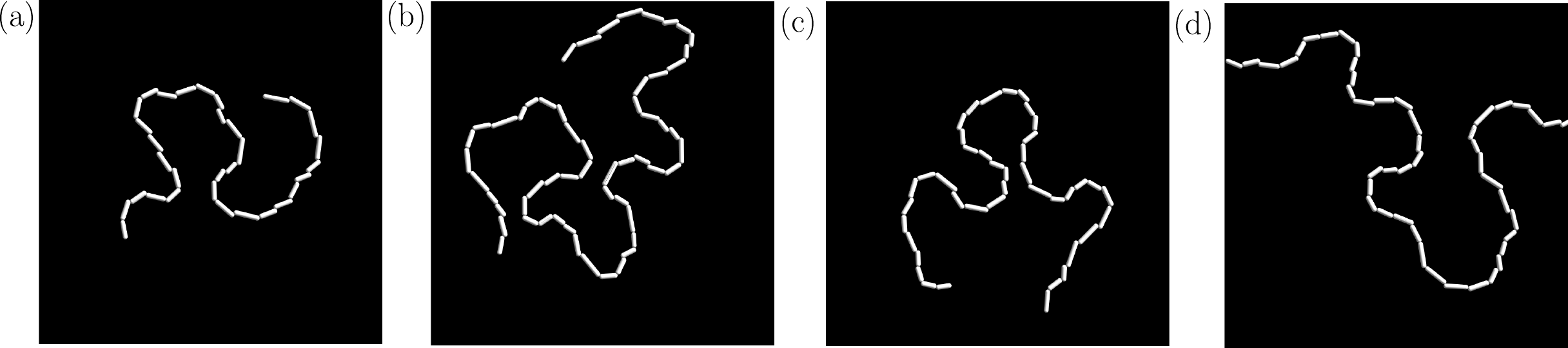}
\caption{\label{fig:trot} Morphology of simulated strings for different values of $t_{\mathrm{rot}}/t_d =  1.33(\mathcal M_i^\theta C_{\mathrm{diff}})^{-1}$: (a) $0.2926$; (b) $0.11704$; (c) $0.05852$; and (d) $0.02926$. The simulations reported in Fig.~\ref{M-fig:simulations} use $t_\mathrm{rot}/t_d = 0.11704$. Varying this parameter does not qualitatively alter the string morphology. The simulations are reported for $U_\mathrm{ee}/k_BT \sim 78$ and $\mathcal{K}\ell_f^2/k_BT \sim 1586$. Note that the above values use $\mathcal M_i^\theta=0.0453$, corresponding to the smallest cell size, just after division.}
\end{figure}

We also considered a nonlinear variant of the fimbrial repulsion, in which the 
fimbriated layer produces a Hertzian-type soft repulsion,
\begin{equation}
    \mathbf F_{ij}^{\mathrm{fm,H}} =
    \begin{cases}
    \mathscr K\,\Lambda_{ij}^{1/2}\left(\Delta_{ij}^{\mathrm{fm}}\right)^{3/2}\hat{\mathbf r}_{ij},
    & \Delta_{ij}^{\mathrm{fm}}>0,\\[6pt]
    \mathbf 0, & \Delta_{ij}^{\mathrm{fm}}\le 0,
    \end{cases}
\end{equation}
with the geometric contact-length prefactor
\begin{equation}
    \Lambda_{ij} = \frac{(l_i+fd)(l_j+fd)}{2(l_i+l_j+2fd)} .
\end{equation}
Since $\Lambda_{ij}^{1/2}(\Delta_{ij}^{\mathrm{fm}})^{3/2}$ has units of area, the 
nonlinear stiffness $\mathscr K$ has units of pressure and is dimensionally 
distinct from the linear stiffness $\mathcal K$. 
Figure~\ref{fig: linear nonlinear} compares colony morphologies obtained with the 
two closures; they are qualitatively similar, and we use the linear form throughout 
the main text for simplicity.

\begin{figure}[H]
\centering
\includegraphics[width=0.5\textwidth]{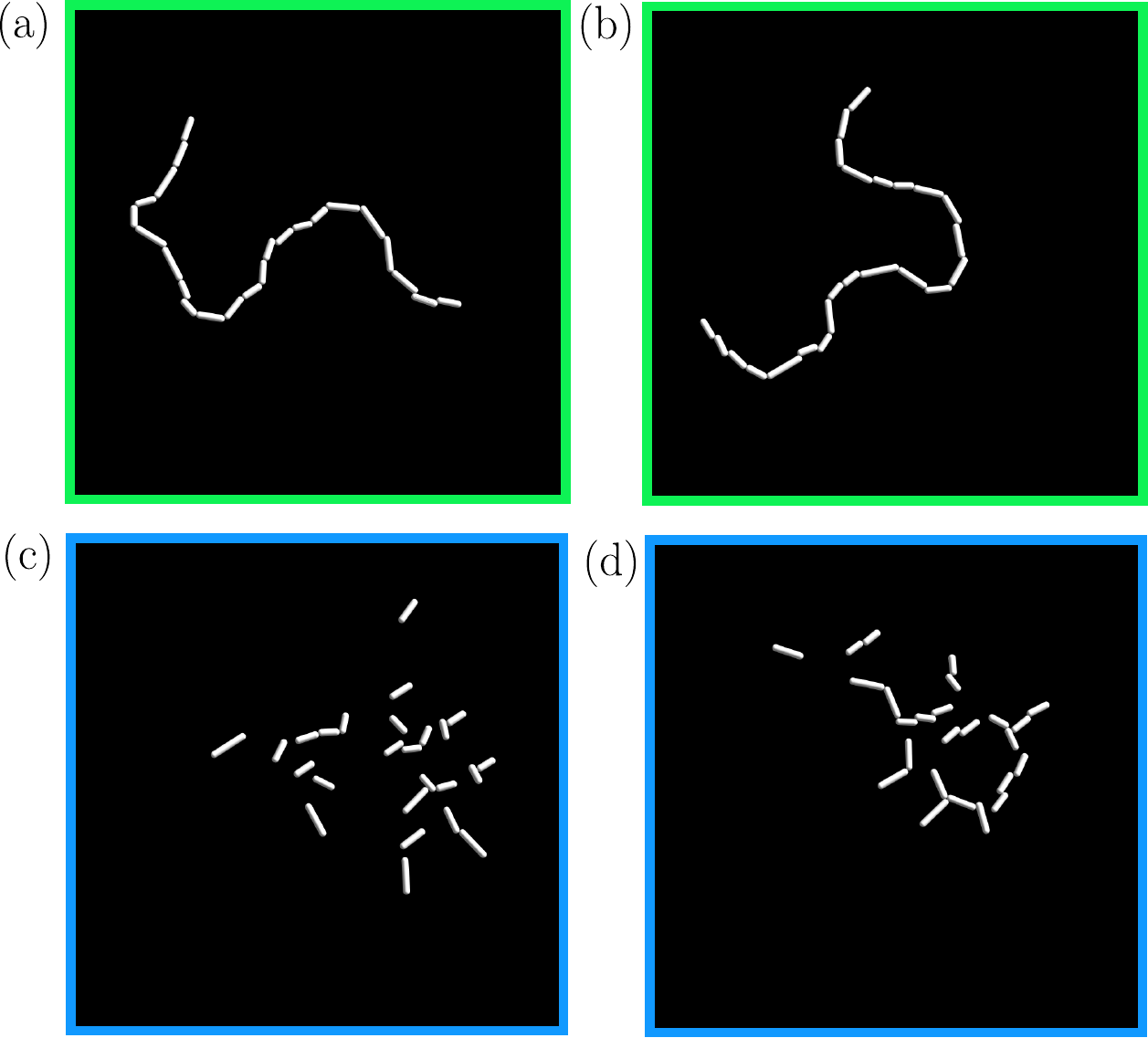}
\caption{\label{fig: linear nonlinear} Comparison of the linear and nonlinear 
fimbrial force models. Colony morphologies after a fixed time interval at 
$U_\mathrm{ee}/k_\mathrm{B}T \sim 80$, for (a) the linear model with 
$\mathcal{K}\ell_{f}^2/k_\mathrm{B}T \sim 1360$ and (b) the nonlinear model with 
$\mathscr{K}\ell_{f}^3/k_\mathrm{B}T \sim 1360$; and at 
$U_\mathrm{ee}/k_\mathrm{B}T \sim$ 3, for (c) the 
linear model with 
$\mathcal{K}\ell_{f}^2/k_\mathrm{B}T \sim 13.6$ and (d) the nonlinear model with 
$\mathscr{K}\ell_{f}^3/k_\mathrm{B}T \sim 13.6$.}
\end{figure}
\par
% Further comments:
% \TD{
% \textit{Consequence of stiffness of bristle layer estimation from beam theory}:  it predicts a value which is orders of magnitude higher than the one used in the simulations to observe string formation. Nevertheless, it supports speculation that after division, the depletion attraction drives a daughter cell to slide along its mother, since doing so monotonically increases the overlap of their excluded volumes; such sliding would sweep the intervening fimbriae tangentially aside rather than compressing them head-on.\\
% \textit{We treat the fimbriated region as a repulsive}, excluded region, however these fimbriae are known to anchor at surfaces using attractive interaction. Maybe we need to specify in detail that the repulsive interaction is only between two fimbriated cells and not true when modelling the fimbriae attachment to a generic solid surface.}

% \begin{figure}[h!]
%     \begin{center}
%     \includegraphics[width=1\textwidth]{fimbriae estimation.pdf}
%     \end{center}
%     \caption{\TD{Experimental estimation of fimbriae density:Uday, can you please update the figure with just the histogram, also if you consider removing the AFM images since they are already in main text?}}
%     \label{fig:AFM images of strains used in this study}
% \end{figure}

%\nocite{*}

\bibliography{bacteria}% Produces the bibliography via BibTeX.